\def\addforarchive{\begin{picture}(0,0)
                   \put(220,210){{\small\sf hep-th/0602051}} 
                   \eP} 

\documentclass{amsproc}

\usepackage{graphicx,amssymb,bbm}
\usepackage{rotating}
\usepackage[mathscr]{eucal}

\copyrightinfo{2006}{J.~Fuchs}

\newtheorem{theorem}{Theorem}

\newtheorem{proposition}[theorem]{Proposition}

\theoremstyle{definition}
\newtheorem{definition}[theorem]{Definition}
\theoremstyle{remark}

\def\alg           {algebra}
\def\auto          {automorphism}
\newcommand\B[3]   {{B_{#1\,#2}}^{\!\!#3}}
\def\be            {\begin{equation}}
\def\bearll        {\begin{array}{ll}}
\newcommand\bigo[1]{\bigoplus_{#1\in\I}}
\def\C             {\ensuremath{\mathcal C}}

\def\cat           {category}
\def\cats          {categories}
\def\cco           {C$_2$-co\-fi\-ni\-te}
\def\cft           {conformal field theory}
\def\cfts          {conformal field theories}
\def\Chi           {{\ensuremath{\mathscr X}}}
\def\chii          {\raisebox{.15em}{$\chi$}}
\def\cir           {\,{\circ}\,}
\def\complex       {{\ensuremath{\mathbbm C}}}

\def\Cong          {\,{\cong}\,}

\def\cva           {conformal vertex algebra}
\def\dim           {{\rm dim}}

\def\diml          {{\rm dim}_{\rm l}}
\def\dimr          {{\rm dim}_{\rm r}}
\def\dS            {diagonalizing S-ma\-trix} 
\def\dsty          {\displaystyle }
\def\ee            {\end{equation}}
\def\eE            {{\rm e}}
\def\eear          {\end{array}}
\def\eP            {\end{picture}}

\def\eq            {\,{=}\,}
\newcommand\erf[1] {(\ref{#1})}
\def\F             {\ensuremath{\mathscr F_{\!0}(\C)}}
\def\findim        {fi\-ni\-te-di\-men\-si\-o\-nal}

\def\ftc           {finite tensor category}
\def\ftcs          {finite tensor categories}
\def\fus           {\,{*}\,}
\def\gz            {generalized }
\def\Hom           {\ensuremath{\mathrm{Hom}}} 
\def\hy            {$\mbox{-\hspace{-.66 mm}-}$\linebreak[0]}
\def\I             {{\mathcal I}}
\def\Ia            {{\mathcal I}^a_{\phantom:}}
\def\id            {\mbox{\sl id}}
\def\idC           {\ensuremath{\mbox{\sl Id}_\C}}
\def\ii            {{\rm i}}

\def\iN            {\,{\in}\,}
\newcommand\inclpic[3] {{\begin{picture}{#3} \scalebox{.#2}
                   {\includegraphics{sydL_#1.eps}}
                   \setlength{\unitlength}{1pt} \eP}}

\def\Ip            {{\mathcal I}'_{\phantom:}}
\def\Is            {{\mathcal I}^*_{\phantom|}}
\def\IsOO          {{\mathcal I}^*_{\phantom:}}
\def\J             {\ensuremath{J}}
\def\jj            {\ensuremath{\jmath}}
\def\kc            {\ensuremath{K_0(\mathcal C)}}

\def\KMKi          {\ensuremath{\widehat{S_i}}}
\def\koe           {\ensuremath{\Bbbk}} 
\def\lea           {{l}}
\def\lwa           {{l{+}1}}
\def\mS            {character S-ma\-trix} 
\def\mtc           {modular tensor category}
\def\mtcs          {modular tensor categories}

\newcommand\N[3]   {{N_{#1\,#2}}^{\!\!#3}}
\def\nE            {\,{\ne}\,}
\def\NN            {{\ensuremath{\mathbb N}}}
\def\normo         {\mbox{\Large\bf:}}
\def\nssi          {non-se\-mi\-sim\-ple}
\def\Nssi          {Non-se\-mi\-sim\-ple}
\def\nua           {{\nu_{\!a}}}
\def\nxt           {\raisebox{.3pt}{$\triangleright$}\hspace*{3pt}}
\def\obj           {{\mathcal O}bj}

\def\one           {{\bf1}}
\def\onematrix     {\mbox{\footnotesize 1$\!$}\!\mbox{\small1}}
\def\epmimo        {$(1{,}p)$~mi\-ni\-mal model}
\def\epmo          {$(1{,}p)$~mo\-del}
\def\oti           {\,{\otimes}\,}
\def\Oti           {{\otimes}}

\def\otiz          {\,{\otimes_\zet^{}}\,}
\def\parfu         {partition function}

\def\P             {\ensuremath{\mathscr P(\C)}}
\def\pf            {\mbox{\sc d}}
\def\PF            {Per\-ron\hy Fro\-be\-ni\-us}

\def\PP            {\ensuremath{Q}}
\def\Pm            {P^-}
\def\Pp            {P^+}
\def\Ppm           {P^\pm}
\def\ps            {{p{-}s}}
\def\q             {quantum }
\def\Q             {Quantum }
\def\qft           {quantum field theory}

\def\qquand        {\qquad{\rm and}\qquad}
\def\RA            {{\mathrm J}}
\def\Rad           {\ensuremath{\mathscr R}}
\def\RB            {\mbox{\j}}
\def\rcva          {rational conformal vertex algebra}

\def\Reg           {\ensuremath{R_{\rm reg}}}
\def\repV          {\ensuremath{\mathcal{R}ep(\V)}}
\def\rep           {repre\-sen\-ta\-ti\-on}
\def\Rep           {{\mathcal R}ep}
\def\rreg          {\ensuremath{\rho_{\rm reg}}}
\def\rmS           {\ensuremath{\mathrm S}}
\def\rS            {ribbon S-ma\-trix} 
\def\rt            {{\ensuremath{\rho_\Tob}}}
\def\rx            {{\ensuremath{\rho_\Chi}}}
\def\rxt           {{\ensuremath{\widehat\rho_\Chi}}}
\def\scs           {\scriptstyle}
\newcommand\setulen[2]{\setlength\unitlength{.#1#2pt}}
\def\slz           {{\rm SL}(2,\zet)}
\def\Sm            {{\ensuremath{S^{\sss\chi_{}}_{\phantom|}}}}
\def\SmO           {\ensuremath{S^{\sss\chi_{}}}}
\def\SmOO          {{\ensuremath{S^{\sss\chi_{}}_{\phantom:}}}}

\def\SO            {S^{\sss\otimes}}
\def\Sd            {{\ensuremath{S^{\sss\otimes}_{\phantom;}}}}
\def\SdO           {S^{\sss\otimes}}
\def\Sdi           {{\ensuremath{S^{\sss\otimes^{\scs -1}}_{\phantom;}}}}
\def\SmS           {\varSigma^{\sss\otimes}}
\def\sr            {{\ensuremath{s^{\dsty\circ\!\!\circ}_{\phantom;}}}}
\def\Sr            {{\ensuremath{S^{\dsty\circ\!\!\circ}_{\phantom;}}}}
\def\srO           {s^{\dsty\circ\!\!\circ}}

\def\ssi           {se\-mi\-sim\-ple}
\def\sss           {\scriptscriptstyle}
\def\ssy           {se\-mi\-sim\-pli\-ci\-ty}

\def\tcs           {tensor categories}

\def\To            {\,{\to}\,}
\def\Tob           {{\ensuremath{\mathscr Y}}}
\newcommand\turlabl[1]{{\begin{turn}{90}$\sss #1$\end{turn}}}
\def\Ump           {U^\mp}
\def\Um            {U^-}
\def\Up            {U^+}
\def\Upm           {U^\pm}
\def\Uqres         {\ensuremath{\overline{\mathrm U_{\mathrm q}}%
                   (\mathfrak{sl}(2))}}
\def\V             {\ensuremath{\mathscr V}}
\def\va            {vertex algebra}

\def\Vee           {{}{}^{\vee\!}}
\def\VEe           {{}{}^{\vee\!\!}}
\def\voa           {vertex operator algebra}
\def\W             {\ensuremath{\mathscr W}}
\def\Y             {\ensuremath{y}}
\def\zet           {{\ensuremath{\mathbb Z}}}
\def\zhuv          {\ensuremath{\mathrm A}(\V)}
\def\zhuvn         {\ensuremath{\mathrm A_n}(\V)}
\def\zzmatrixabcd  {\mbox{$\Big(\!\begin{array}{cc}\!\scs a\!\!\!&\!\!\scs b\!%
                    \\[-3pt]\!\scs c\!\!\!&\!\!\scs d\!\eear\!\Big)$}}
\def\zzmatrixS     {\mbox{$\Big(\!\begin{array}{cc}\!\scs 0\!\!\!&\!\!\scs-1\!%
                    \\[-3pt]\!\scs 1\!\!\!&\!\!\scs 0\!\eear\!\Big)$}}
\def\zzmatrixT     {\mbox{$\Big(\!\begin{array}{cc}\!\scs 1\!\!\!&\!\!\scs 1\!%
                    \\[-3pt]\!\scs 1\!\!\!&\!\!\scs 0\!\eear\!\Big)$}}

\begin{document}

\title[Non-semisimple fusion rules]
        {On non-semisimple fusion rules and tensor categories} 
\author
       {J\"urgen Fuchs}
\address{Avdelning fysik, Karlstads Universitet,
       Universitetsgatan~5, S\,--\,65188\, Karlstad}
\email{jfuchs@fuchs.tekn.kau.se}

\thanks{This work is supported by VR under project no.\ 621--2003--2385.}

\subjclass[2000]{81T40,
                 18D10,
                 19A99,
                 17B69}
\date{}   

\begin{abstract}
Category theoretic aspects of non-rational conformal field theories are 
discussed. We consider the case that the category \C\ of chiral sectors is a 
finite tensor category, i.e.\ a rigid monoidal category whose class of objects 
has certain finiteness properties. Besides the simple objects, the 
indecomposable projective objects of \C\ are of particular interest.
\\
The fusion rules of \C\ can be block-diagonalized. A conjectural connection 
between the block-diagonalization and modular transformations of characters 
of modules over vertex algebras is exemplified with the case of the 
$(1{,}p)$ minimal models.
         \addforarchive
\end{abstract}
\maketitle


\section*{Introduction}

Rational two-dimensional conformal field theory (rational CFT, or RCFT, for 
short), corresponding to vertex algebras with \ssi\ \rep\ \cat, is a 
well-developed subject. In contrast, in the \nssi\ case our knowledge is 
considerably more limited. It is, for instance, not even clear which sub\cat\ 
of the \cat\ of all generalized modules of a conformal vertex algebra can play 
the role of the \cat\ \C\ of chiral sectors of an associated \cft. In this note 
we discuss a few features of the \nssi\ case, in particular the relevance of 
indecomposable projective objects of \C, as well as a relationship between the 
Grothendieck ring of \C\ (which, by the assumed properties of \C, exists) and 
modular transformations of characters, which is conjectured to hold for a 
certain class of non-rational CFTs. We also recall some aspects of
the Verlinde relation, to which the conjecture reduces in the \ssi\ case.
  
Crucial input for the study of chiral \cft\ comes from the work by Jim Lepowsky
and his collaborators on vertex 
algebras and their \rep s. It is encouraging that the tensor product 
theory for modules over vertex algebras initiated in \cite{hule3.5} is being 
developed further (see \cite{miya10,hulz}) so as to cover a more general class 
of modules. One should indeed be confident that these efforts will eventually 
allow for an enormous improvement of our understanding of non-rational \cfts.


\section{Fusion rules and rational vertex algebras}\label{sec.i}

In rational CFT the rank of the 
sheaf of conformal $p$-point blocks on a Riemann surface of genus $g$ is given 
by
  $$
  N^{\sss(g)}_{i_1i_2...i_p} = \sum_{m\in\I} \big(S_{0,m}^{}\big)^{2-2g}_{}
  \prod_{a=1}^p \frac{S_{i_a, m}^{}} {S_{0,m}^{}} \,,
  $$
where $\I$ labels the set of chiral sectors of the theory, with $0\iN\I$ 
corresponding to the vacuum sector, and $S$ a unitary $|\I|{\times}|\I|$-matrix 
determined by the theory. This formula is known as the {\em Verlinde formula\/},
though sometimes this term is reserved for the particular case $p\eq0$ (i.e., 
the sheaf of genus-$g$ characters), or for genus 0 and $p\eq 3$. 
In the latter case, 
  \be
  \N ijk \equiv N^{\sss(0)}_{i\,j\,k^\vee}
  = \sum_{m\in\I} \frac{S_{i,m}^{}\,S_{j,m}^{}\,S_{m,k}^{-1}} {S_{0,m}^{}}
  \label{v} \ee
are the {\em fusion rules\/} of the RCFT. If the conformal blocks possess 
appropriate factorization properties, then the formula for 
$N^{\sss(g)}_{i_1i_2...i_p}$ is actually implied by \erf v. Heuristic arguments 
for the validity of \erf v for any RCFT were already given when this equality 
was discovered for a particular class of such theories, the $\mathfrak{sl}(2)
_\ell$ WZW \cfts\ at positive integral level $\ell$ \cite{verl2}.

When stating this conjecture one faces the problem that apparently its very
formulation presupposes a thorough understanding of (rational) chiral 
CFT, including e.g.\ an accurate definition of conformal blocks. Fortunately, 
however, strictly less information is needed: One can formulate, and prove, the 
conjecture entirely in terms of the \rep\ theory of the chiral symmetry 
algebra of the CFT. The latter is\,%
  \footnote{~%
  In another approach to conformal \qft, the role of the chiral symmetry algebra
 is taken over by a net of von Neumann algebras, see e.g.\ \cite{boek1,rehr22}.}
a \cva, and we call a CFT rational iff this \cva\ is rational. This is still not
unequivocal, because several similar, but inequivalent, interpretations of the 
qualification `rational' for a \va\ are in use.  We resolve this remaining 
vagueness by adopting the axioms used e.g.\ in theorem 5.1 of \cite{huan24}:\,%
\begin{definition}
\label{def.r}
A \cva\ \V\ is called {\em rational\/} iff it obeys the following conditions:
\\[1pt]
\nxt~\V\ is simple; 
\\[1pt]
\nxt~$\V^{(0)}\Cong\complex\,$ and \,$\V^{(n)}\eq0$ for $n\,{<}\,0$\,; 
\\[1pt]
\nxt~~\V\ is \cco; 
\\[1pt]
\nxt~every $\mathbb{N}$-gradable weak \V-module is fully reducible;
\\[1pt]
\nxt~every simple \V-module not isomorphic to \V\ has positive conformal 
weight.
\end{definition}

Two key properties of \rcva s make a purely \rep\ theoretic formulation of the
Verlinde conjecture possible. The first is based on the notion of character; the
character $\chii_W$ of a \V-mo\-du\-le $W$ is the graded dimension
  \begin{equation}
  \chii_W(q) = {\rm tr}_{W}^{} \Big( q^{L_0-c/24}\,Y_{\!W}(v_\Omega^{}) \Big)
  = \sum_{n\ge0} \mbox{dim}(W^{(n)})\,q^{n+\Delta_i-c/24} \,,
  \label{chii} \ee
and there are analogous trace functions in which the vacuum vector $v_\Omega$
is replaced by an arbitrary vector $v\iN \V$.
These can be analytically continued from the region $0\,{<}\,|q|\,{<}\,1$, 
in which they are absolutely convergent, to
analytic functions of $q\eq\eE^{2\pi\ii\tau}$ for $\tau$ in the complex upper 
half-plane. Now a rational \cva\ has, up to isomorphism, only finitely many 
irreducible \V-modules $U_i$, $i\iN\I$. Further, as shown in \cite{dolm13} and 
(with slightly different assumptions than those of definition \erf{def.r}) in 
\cite{zhu3}, for any $v\iN \V$ the \findim\ vector space of trace functions, 
and in particular the space $\Chi\eq{\rm span}\{\chii_{U_i}\,|\,i\iN\I\}$ of 
characters, carries a \rep\ \rx\ of the modular group \slz: 
  $$
  \slz \ni \zzmatrixabcd = \gamma \mapsto \rx(\gamma) \,, \qquad
  \chii_{U_i}(\gamma\tau) = \sum_{j\in\I}\rx(\gamma)_{ij}\,\chii_{U_j}(\tau) \,,
  $$
where $\gamma\tau\eq\frac{a\tau+b}{c\tau+d}$. This allows one to define the 
right hand side of the relation \erf v: $S$ is the matrix
  \be
  \Sm = \rx(\zzmatrixS)
  \label{Sm}\ee
that represents the modular S-transformation $\tau\,{\mapsto}\,{-}\frac1\tau$ 
on the characters. This matrix is commonly referred to as the {\em modular\/} 
S-matrix, but to ensure that it is not mixed up with one of the matrices \Sd\ 
and \Sr\ to be introduced below, we shall henceforth call it the {\em \mS\/} 
instead.

The second key ingredient is the tensor product of \V-modules and of 
intertwiners (see e.g.\ \cite{hule3.5,hule3,lI0,huan3}), which endows the \rep\ 
\cat\ \repV\ with the structure of a braided monoidal \cat. For rational \V, 
this \cat\ has further special properties. Indeed, as 
shown in \cite{huan21,huan24}, one has

\begin{proposition}
The \rep\ \cat\ of a rational \cva\ is a \mtc.
\end{proposition}

Recall that a \mtc\ is the following structure (see \cite{TUra}, where some of 
the conditions imposed below are slightly relaxed, and e.g.\ \cite{fuRs4}):

\begin{definition}
A {\em \mtc\/} \C\ is a \cat\ with the following properties:
\\[1pt]
\nxt~\C\ is abelian, \complex-linear and \ssi;
\\[1pt]
\nxt~\C\ is monoidal, with simple tensor unit;
\\[1pt]
\nxt~\C\ is a {\em ribbon\/} \cat;
\\[1pt]
\nxt~the number of isomorphism classes of simple objects of \C\ is finite;
\\[1pt]
\nxt~the braiding is maximally non-degenerate, in the sense that the 
matrix $\srO$ with entries 
  \begin{equation}
  \srO_{i,j} \,{:=}\, (d_{U_j}^{}\oti\tilde d_{U_i}^{}) \cir
  [\id_{U_j^\vee}\oti(c_{U_i,U_j}^{}{\circ}\,c_{U_j,U_i}^{})\oti\id_{U_i^\vee}]
  \cir (\tilde b_{U_j}^{}\oti b_{U_i}^{})
  \label{sr} \ee
in $\Hom(\one,\one)\Cong\complex$ is invertible.
\end{definition}

In \erf{sr}, $i$ and $j$ take values in the set $\I$ labeling the isomorphism
classes of simple objects of \C, and $\{U_i\}$ is a set of representatives for 
those classes, such that $U_0\eq\one$ is the tensor unit.
That \C\ is ribbon\,%
  \footnote{~%
  Besides the qualifier `ribbon' \cite{retu2}, which emphasizes the
  similarity with the properties of ribbons in a three-manifold, also the
  terms `tortile' \cite{joSt6,shum} and `balanced rigid braided' are in use.}
means that there are families $\{c_{U,V}\}$ of braiding isomorphisms
in $\Hom(U\oti V,V\oti U)$, $\{\theta_U\}$ of balancing twist isomorphisms
in $\Hom(U,U)$, $\{b_U\}$ of coevaluation evaluation morphisms in 
$\Hom(\one,U\oti U^\vee)$, and $\{d_U\}$ of evaluation morphisms in 
$\Hom(U^\vee{\otimes}\,U,\one)$, respectively,
with appropriate properties; see, for instance, chapter 2 of \cite{BAki}.
Every ribbon \cat\ is {\em rigid\/}, i.e.\
besides the right duality given by $\{d_U,b_U\}$ there is
also a left duality (with coevaluation and evaluation morphisms to be denoted
by $\{\tilde b_U,\tilde d_U\}$), and it is even {\em sovereign\/}, i.e.\ the 
left and right duality functors are equal, ${{}^\vee\!}U\eq U^\vee$ for all 
$U\iN\obj(\C)$ and ${{}^\vee\!}\!f\eq f^\vee$ for all morphisms of \C.

In a \mtc\ the tensor product induces a multiplicative ring structure on the 
Grothendieck group \kc, i.e.\ the quotient of the free abelian group generated 
by isomorphism classes $[U]$ by the ideal generated by the relations 
$[U] \eq [V]\,{+}\,[W]$ for all exact sequences $0\To V\To U\To W\To 0$. 
We denote by $\N ijk\iN\zet_{\ge0}$ the structure constants of the ring \kc\
in the basis furnished by the classes $\{[U_i]\,|\,i\iN\I\}$ of simple objects.
These numbers are the fusion rules that appear on the left hand side of \erf v.

\smallskip

Thus in short, for an RCFT with chiral symmetry \V\ the Verlinde formula \erf v
expresses the structure constants of the based ring $K_0(\repV)$ in terms 
of the \mS\ \erf{Sm} for the characters of \repV-\rep s.

On the other hand, an expression for $\N ijk$ of the form \erf v also follows
directly from the \rep\ theory of the {\em fusion algebra\/}
$K_0(\repV) \otiz \complex$. Indeed, there is a (not uniquely 
determined) {\em diagonalizing S-matrix\/} $\Sd$ such that \erf v holds with 
$S\eq\Sd$ (see section \ref{sec.ssifrv} below for details). 
Thus the contents of \erf v may also be summarized by asserting the equality
  \begin{equation}
  \Sd = \Sm
  \label{SdSm} \ee
between (a possible choice of) the \dS\ \Sd\ and the \mS\ \Sm. 

Moreover, as already hinted at by the choice of symbol $s$ in \erf{sr}, for a
\mtc\ there is also a third S-matrix of interest, namely the matrix
  \begin{equation}
  \Sr = \zeta\,\sr \qquad{\rm with}\quad\
  \zeta := \big( \mbox{\large$\sum_{i\in\I}$} (\srO_{i,0})^2_{} \big)^{-1/2}_{}
  \label{Sr} \ee
(one can show that $\srO_{i,0}\,{\ge}\,0$), which is defined in terms of the 
braiding and the dualities of \repV, and which we will accordingly call the 
{\em \rS\/}. And indeed, for the \rep\ \cat\ of any \rcva\ in addition to 
\erf{SdSm} one also has the equality
  \begin{equation}
  \Sr = \Sd 
  \label{SrSd} \ee
between the \rS\ and the \dS. Note that, unlike \erf{SdSm}, this is a 
statement about the \mtc\ \C\ and can thus be formulated without reference 
to the underlying vertex \alg\ that has \C\ as its \rep\ \cat. 

Both of the equalities \erf{SdSm} and \erf{SrSd} have been proven in the
adequate contexts: \erf{SrSd} for \mtcs, and \erf{SdSm} for \rcva s.
We will present some details, including a proof of \erf{SrSd}, in section 
\ref{sec.ssifrv} below. But in this paper our main interest is in a broader 
class of CFTs which are not necessarily rational. For these theories a priori
neither of the matrices \SmOO\ and \Sd\ exists any longer. However, there is 
at least one interesting class of models, the \epmimo s \cite{kaus},
for which such matrices can indeed be extracted from \repV, such that 
by postulating the validity of \erf{SdSm}, the corresponding generalization of 
the Verlinde formula yields sensible fusion rules. Like in the rational case, 
an understanding of this relationship requires a good grasp
both on purely categorical and on \rep\ theoretic aspects of the theory.
Here we concentrate on categorical aspects. On the \rep\ theoretic side,
the crucial new ingredient is the generalization of the tensor product theory 
for \repV\ \cite{hule3.5,hule3,huan3} to include \gz \V-modules and logarithmic
intertwining operators that has been presented in \cite{miya10,hulz}.
 
\medskip

The rest of this note is organized as follows. 
We start in section \ref{sec.bftcs} by discussing a class of \tcs\ that 
includes, besides all \mtcs, the \rep\ \cats\ for the \epmimo s that were 
suggested in the literature, and which we expect to encompass many more 
non-rational CFT models of interest. Afterwards we turn our attention to 
properties of the ($K_0$-)\,fusion rules of such \cats, illustrating them 
first in the \ssi\ case in section \ref{sec.ssifrv}, and then discussing the 
general case in section \ref{sec.nssi}. Next we comment, in section 
\ref{sec.mt}, on the modular transformations of characters, which in the 
\nssi\ case do not span the space of conformal zero-point blocks on the torus. 
In section \ref{sec.epmimo} pertinent information on the \epmimo s is 
collected, including in particular, in section \ref{sec.epmimo}.3, the 
conjecture of \cite{fhst} for the fusion rules of the $(1{,}p)$ models. 
In the final section \ref{sec.o} we remark on the significance of some 
peculiarities of these $(1{,}p)$ fusion rules, and also point out that taking 
the relation between chiral and full CFT into consideration may allow one to 
gain additional insight into the purely chiral issues addressed here.


\section{Braided finite \tcs}\label{sec.bftcs}

Eventually one would like to understand the vertex \alg\ and its \rep\ theory 
for any arbitrary \cft. But in view of the limited information available, it 
would be presumptuous to attack all models at once. As a modest first step one 
can instead try to learn more about those non-rational CFT models which are 
the closest relatives of the rational ones. To us, the crucial properties 
seem to be that up to isomorphism there is still a finite number of 
irreducible \V-modules and that all modules have a Jordan-H\"older series of 
finite length. Unfortunately, even at this restricted level of ambition it is 
far from obvious how to characterize the relevant \cats. Certainly at the \rep\ 
theoretic level one has to deal with {\em weak\/} or {\em generalized modules\/} 
\cite{mila3,miya10,hulz}, 
for which, unlike for ordinary \V-modules, semisimplicity of the action of the
Virasoro zero mode $L_0$ is not required. Further, the
relevant subcategory \C\ of the \cat\ of all \gz \V-modules must still admit
   \def\leftmargini{1.1em}~\\[-1.2em]
   \begin{itemize}\addtolength\itemsep{2pt}
\item[$\triangleright$] 
   a tensor product, as the \rep\ theoretic foundation of operator product 
   expansions;
\item[$\triangleright$] 
   a braiding, to account for monodromy properties of conformal blocks; and 
\item[$\triangleright$] 
   a notion of contragredient \rep\ \cite{frhl}, in order to allow the 
   collection of two-point blocks on the Riemann sphere to be nondegenerate.
\end{itemize}

Sub\cats\ \C\ with these properties certainly exist \cite{hulz}, 
though it can be difficult to verify for concrete models that a specified class
of modules of, say, a \W-algebra in \cft\ for which the associated vertex \alg\
is not known in full detail, satisfies the relevant criteria. If \V\ is \cco,
then the class of finite-length weak \V-modules has the desired properties
\cite[Theorems\,4.5\,\&\,5.4]{miya10}. In any case, \C\ should contain enough 
projectives, and in particular contain all simple \V-modules as well as their 
projective covers. (For \cco\ vertex algebras, the latter is again fulfilled 
for the \gz\ modules of finite length \cite[Theorem\,6.4]{miya10}.) 
Accordingly, for the specific 
class of models we have in mind, the structure of a {\em \ftc\/} \cite{etos} 
seems to provide a suitable framework. Recall from \cite{etos} the following

\begin{definition}
Let \C\ be an abelian \cat\ with \alg ically closed ground field \koe\ and
\findim\ morphism spaces.
\\[.2em]
(i)\,\, \C\ is called {\em finite\/} iff the number of isomorphism classes of 
simple objects is finite, every simple object has a projective cover, and 
\C\ is artinian, i.e.\ every object has finite length. 
\\[.2em]
(ii) A {\em \ftc\/} is a finite rigid monoidal \cat\ with simple tensor unit.
\end{definition}

{}From now on \C\ will stand for a \ftc\ (in the application to \cft, \koe\
is \complex). We continue to write $\{U_i\,|\,i\iN\I\}$ for a (finite) set of 
representatives of isomorphism classes of simple objects of \C, with $U_0\eq\one
$. For $i\iN\I$ we denote by $i^\vee\iN\I$ and $\Vee i\iN\I$ the unique labels 
such that $U_i^\vee\Cong U_{i^\vee}$ and $\Vee U_i\Cong U_{\Vee i}$, 
respectively. Besides the simple objects, the indecomposable projective 
objects turn out to be particularly important. We denote by 
$\{P_i\,|\,i\iN\I\}$ the set of indecomposable projective covers of the
simple objects $U_i$.

Let us list some properties of \ftcs:
\\[2pt]
\nxt~%
The rigidity of \C\ guarantees \cite[Proposition\,2.1.8]{BAki}
that the tensor product bifunctor is exact in both of its arguments.
\\[2pt]
\nxt~%
By rigidity,
  \begin{equation}
  \Hom(U\oti V,W) \cong \Hom(V,\Vee U\Oti\,W) \cong \Hom(U,W\oti V^\vee) \,.
  \label{hhh} \ee
While the left and right dualities need not coincide, still
$\VEe L\Cong L^{\!\vee}$ for every invertible object $L$.
\\[2pt]
\nxt~%
A simple object $U$ is also absolutely simple, i.e.\ satisfies
$\Hom(U,U)\eq\koe\,\id_U$.
\\[2pt]
\nxt~%
For all $i\iN\I$, $\Hom(P_i\oti\Vee U_i,\one)\nE 0$ and
$\Hom(P_i\oti\Vee U_i,P_0)\nE 0$ \cite[Remark\,in\,2.7]{etos}.  
\\[2pt]
\nxt~%
For any object $U$ and any projective object $P$, the objects $P\oti U$ and 
$U\oti P$ are projective \cite[IV,\,Corollary\,2\,on\,p.\,441]{kaluX}.
In particular, after tensoring with a projective object every exact sequence 
splits.
\\[2pt]
\nxt~%
As a consequence \cite[Proposition\,2.3]{etos}, together with $P$ also $P^\vee$ 
is projective. This, in turn, implies that any indecomposable projective object 
has, up to isomorphism, a unique simple subobject, and that any projective
object is also injective, and vice versa.
\\[2pt]
\nxt~%
It also follows that if $\one$ is projective, then, owing to $U\Cong U\oti\one$,
so is every object $U$ of \C. This implies that the tensor unit is projective 
iff \C\ is \ssi.
\\[2pt]
\nxt~%
Further, the assignment
  $$
  i \mapsto i^+ \qquad{\rm such~that}\quad P_i^\vee \Cong P_{i^+} 
  $$
constitutes a bijection from $\I$ to itself, satisfying $i^{++}\eq i^{\vee\vee}$.

\smallskip

Next we note that, being exact in both arguments, the tensor product bifunctor 
induces a ring structure on \kc. We call this structure the fusion product and 
denote it by $[U]\fus[V] \,{:=}$\linebreak[0]$[U\Oti V]$,
and refer to the \koe-algebra
  \begin{equation}
  \F := \kc \otiz \koe
  \label{F} \ee
as the {\em fusion algebra\/} of \C.
For any object $U$, the matrix of left (or right) multiplication by $[U]$ in 
\kc\ has nonnegative entries; the \PF\ eigenvalue of this matrix is called 
the \PF\ dimension of $U$.

Besides its ring structure, \kc\ inherits further properties from \C:
Since the tensor product is associative up to isomorphism, the fusion product 
$\fus$ is associative. Similarly, the class $[U_0]$ of the tensor unit is the 
unit element of \kc, $[U_0]\fus[U]\eq[U]\eq[U]\fus[U_0]$, while rigidity implies
that $[U^\vee]\fus[U]\eq[U_0]+\ldots$ and $[U]\fus[\Vee U]\eq[U_0]+\ldots$\,.\,%
  \footnote{~%
  But note that for $[U_0]$ to occur in the decomposition of $[U]\fus[V]$ with
  respect to the fusion basis it suffices that $U\oti V$ has the tensor unit as
  a subquotient, which can happen even for simple objects $U$, $V$ that are not
  isomorphic to each other's left or right duals.}
And owing to finiteness of \C, the algebra
\F\ has a \zet-basis $\{[U_i]\}_{i\iN\I}^{}$, to be called the 
{\em fusion basis\/}, in which the structure constants are nonnegative integers. 
As in the case of \mtcs, we write $\N ijk$ for these numbers, i.e. write 
  $$
  [U_i]\fus[U_j] = \sum_{i\in\I} \N ijk\, [U_k] \,,
  $$
and refer to the integers $\N ijk$ as the ($K_0$-)\,{\em fusion rules\/} of \C. 

In terms of these fusion rules, the tensor products of the simple objects
$U_i$ and indecomposable projective objects $P_i$ are given by
\cite[Proposition\,2.2]{etos}
  $$
  P_i \oti U_j \cong \bigo k \N k{j^\vee}i\, P_k \qquand
  U_j \oti P_i \cong \bigo k \N {\Vee j}ki\, P_k \,.
  $$
The objects $P_i$ close among themselves under the tensor product, in
the sense that 
  \begin{equation}
  P_i \oti P_j \cong \bigo k \B ijk\, P_k \qquad{\rm with}\quad
  \B ijk\iN\zet_{\ge0} 
  \label{B} \ee
for $i,j\iN\I$. 

The bijection $i\,{\mapsto}\,i^+$ satisfies $\N{0^+}{j^\vee}i\eq\delta_{i^+,j}$, 
implying that the object $U_{0^+}$ is invertible and obeys
  $$
  P_{i^+} \cong P_{\Vee i} \oti U_{0^+} \,, \qquad
  U_{i^+} \cong U_{\Vee i} \oti U_{0^+} \,, \qquad
  P_{i^{+\vee}} \cong U_{0^{+\vee}} \oti P_i 
  $$
as well as
  $$
  P_{i^{\vee\vee}} \cong U_{0^{+\vee}} \oti P_{\Vee\Vee i} \oti U_{0^+} 
  $$
\cite[Lemmata\,2.9,2.10]{etos}.
  
\medskip

Braid group statistics is a characteristic feature of low-dimensional
quantum field teories. Accordingly
we now assume that \C\ is a {\em braided\/} \ftc. (Braided \tcs\ with similar
properties as those of \ftcs\ have also been studied in \cite{lyub7,KElu}.)
Then the ring \kc\ is
commutative, and the isomorphisms \erf{hhh} are supplemented by
  $$
  \Hom(U\oti V,W) \cong \Hom(V,U^\vee\Oti\,W) \cong \Hom(U,W\,\Oti\Vee V) \,.
  $$
Also (compare e.g.\ formula (2.2.6) of \cite{BAki}), the double dual functors 
$\mbox{\sl?}^{\vee\vee}$ and ${}^{\vee\vee\!}\mbox{\sl?}$ are naturally 
isomorphic (as functors, though in general not as monoidal functors) to the
identity functor \idC. One can thus define left and right (quantum)
dimensions $\diml(U),\dimr(U)\iN\Hom(\one,\one)$ of objects of \C\ by
  $$
  \diml(U) := \tilde d_U \cir c_{\Vee U,U}^{} \cir \tilde b_U \,, \qquad
  \dimr(U) := d_U \cir c_{U,U^\vee}^{} \cir b_U \,.
  $$ 
By (absolute) simplicity of $\one$, when $\dimr(U)\iN\Hom(\one,\one)\Cong\koe$ 
(or $\diml(U)$) is non-zero, then it is an isomorphism and $\one$ is a retract 
of the object $U\oti U^\vee$ (respectively, of $\Vee U\oti U$). It follows
(compare \cite[Theorem\,2.16]{etos}) that, unless \C\ is \ssi,
$\diml(P)\eq0\eq\dimr(P)$ for every projective object $P$.


\section{Semisimple fusion rules and the Verlinde formula}\label{sec.ssifrv}

In this section we assume that \C\ is a {\em \ssi\ sovereign\/} braided \ftc,
e.g.\ a \mtc.
Then in particular every object is projective, and \erf B reduces to
  $$
  U_i \oti U_j \cong \bigo k \N ijk\, U_k 
  $$
for $i,j\iN\I$. 

Combining \erf{hhh} with \ssy,
one has $\Hom(U^\vee\Oti\,V,W)\Cong\Hom(U\oti W,V)$, and thus the \rep\
matrix $\rreg([U^\vee])$ of $[U^\vee]$ in the (left, or equivalently right) 
regular \rep\ \rreg\ of the fusion algebra \F\ is the transpose of $\rreg([U])$.
Commutativity of \F\ then implies that all matrices $\rreg(\,\cdot\,)$ are normal 
and commute with each other, and therefore can be diagonalized simultaneously. 
Thus \rreg\ is fully reducible, and hence \F\ is {\em \ssi\/}. 

\smallskip

So \F\ is a \ssi\ unital commutative associative \alg\ over an \alg ically 
closed field \koe. The structure of such algebras is easily worked out.\,%
  \footnote{~For a few more details about \ssi\ fusion \alg s see
  e.g.\ \cite{kawA,jf24,gann16}.}
First, there is a basis $\{e_l\,|\,l\iN\Is\}$ of idempotents, satisfying
  $$ 
  e_l^{}\fus e_m = \delta_{lm}\,e_m  \qquand
  \mbox{\large$\sum_{l\in\Is}$}\, e_l^{} = e \,,
  $$
such that there is an isomorphism
  $$
  \F \,\cong\, \bigoplus_{l\in\Is} \koe e_l^{} 
  $$
of \koe-\alg s. Note that the label sets $\I$ and $\Is$ are in bijection, but 
there does not exist any natural bijection between them.

The basis transformation $[U_i]\eq\sum_{l\in\IsOO}\!\PP_{i,l}\,e_l$, $i\iN\I$, 
from the basis of idempotents to the fusion basis defines an invertible 
matrix \PP, unitary up to normalisation, in terms of which the structure 
constants can be written as $\N ijk\eq\sum_l \PP^{}_{i,l}\, \PP^{}_{j,l}\, 
\PP^{-1}_{l,k}$. The matrix \Sd\ with entries
  \begin{equation}
  \SdO_{i,l} := \xi_l\, \PP_{i,l} \qquad {\rm with}\quad\ 
  \xi_l:= \big(\sum_{i\in\I}|\PP_{il}|^2_{}{\big)}^{\!-1/2}_{}\, {=}\,\,
  \SdO_{0,l} \ne 0
  \label{Sd} \ee
is unitary and diagonalizes the fusion rules,
  \begin{equation}
  \N ijk = \sum_{l\in\Is}\frac{\SdO_{i,l}\,\SdO_{i,j}\,
  \Sdi\!\!\!\!\!\!\!{}_{l,k}^{}} {\SdO_{0,l}} \,.
  \label{Nd} \ee
It is thus the diagonalizing matrix announced before formula \erf{SdSm} in
section \ref{sec.i}. 

\medskip

We are now in a position to establish the postulated equality \erf{SrSd}.
One possibility is the following.
After tacitly replacing \C\ with an equivalent strict monoidal \cat, we can
employ the usual graphical notation (see e.g.\ \cite{BAki,fuRs4} and the
literature cited there)
\\
  \inclpic{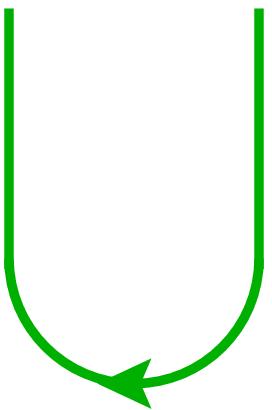}{3} {(170,40)(-80,0) \put(-28,15){$b_{U}^{}\,{=}$} \setulen75
                  \put(3,38){\turlabl U} \put(32,38){\turlabl{U^{\!\vee}}} }
  \inclpic{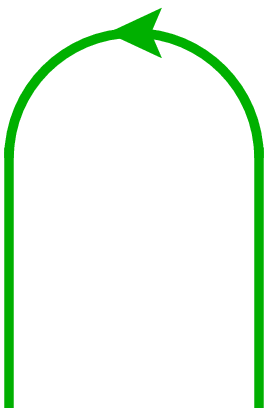}{3} {(95,40)(0,0) \put(-28,15){$d_{U}^{}\,{=}$} \setulen75
                  \put(3,0){\turlabl{U^{\!\vee}}} \put(32,0){\turlabl U} }
  \inclpic{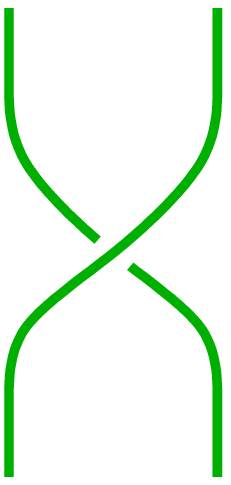}{3} {(10,40)(0,3) \put(-34,18){$c_{U,V}^{}\,{=}$}
                  \setulen75 \put(4,0){\turlabl U} \put(28,0){\turlabl V} }
\\[6pt]
for the (co)evaluation and braiding morphisms.
Then the following chain of equalities is easy to verify:

\begin{picture}(0,121)(23,-66)
\setlength\unitlength{1.2pt}
    \put(5,17)    {$\dsty\frac{\srO_{i,k}}{\srO_{0,k}}\,\srO_{j,k}\hspace*{.5em}
                    =\hspace*{.5em}\dsty\frac{\srO_{i,k}}{\srO_{0,k}}$}
    \put(76,0)     {\scalebox{.15}{\includegraphics{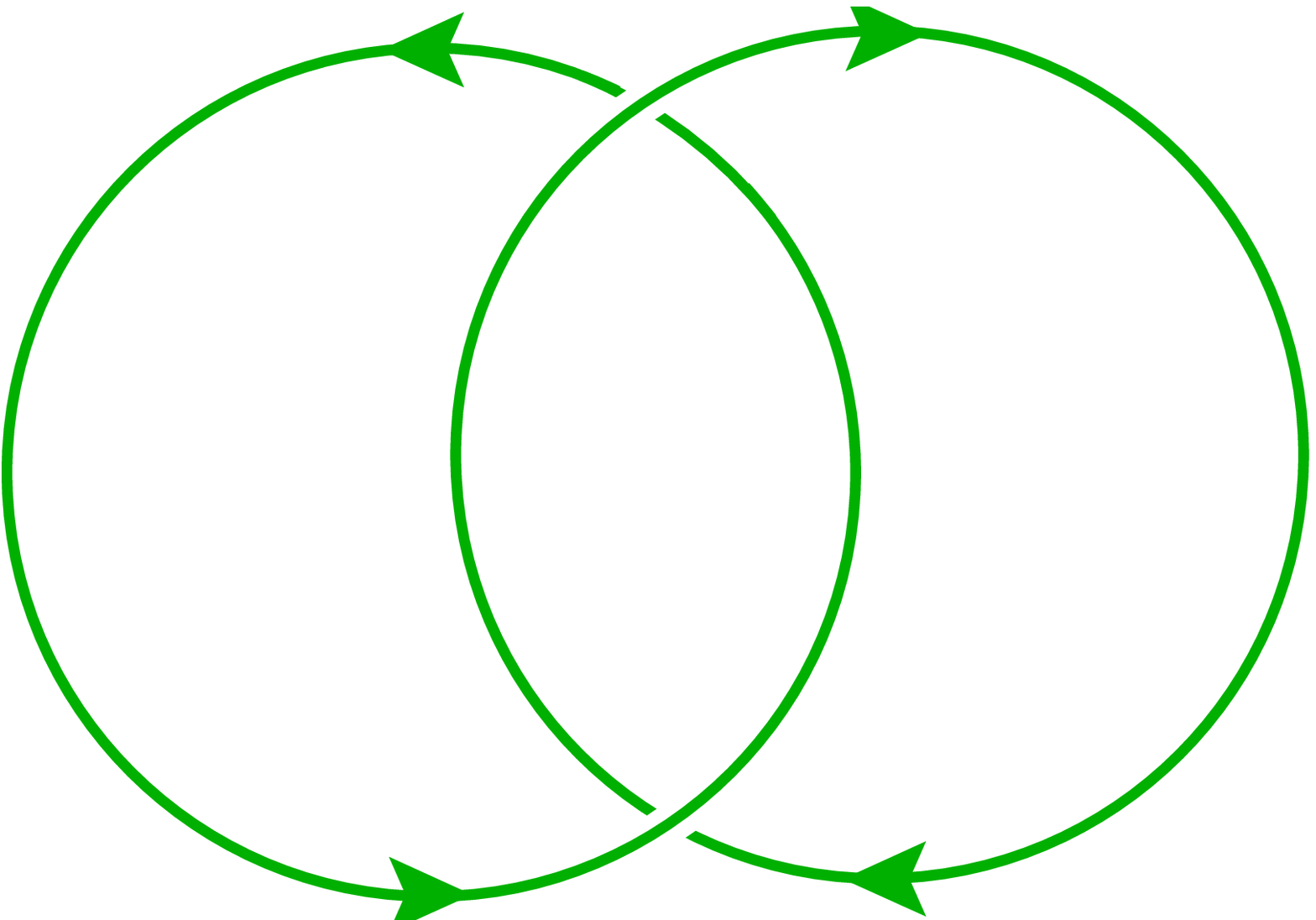}}}
    \put(92.3,17.5){$\sss j$}
    \put(115.5,17.5){$\sss k$}
    \put(144,17)   {$=$}
    \put(163,0)    {\scalebox{.15}{\includegraphics{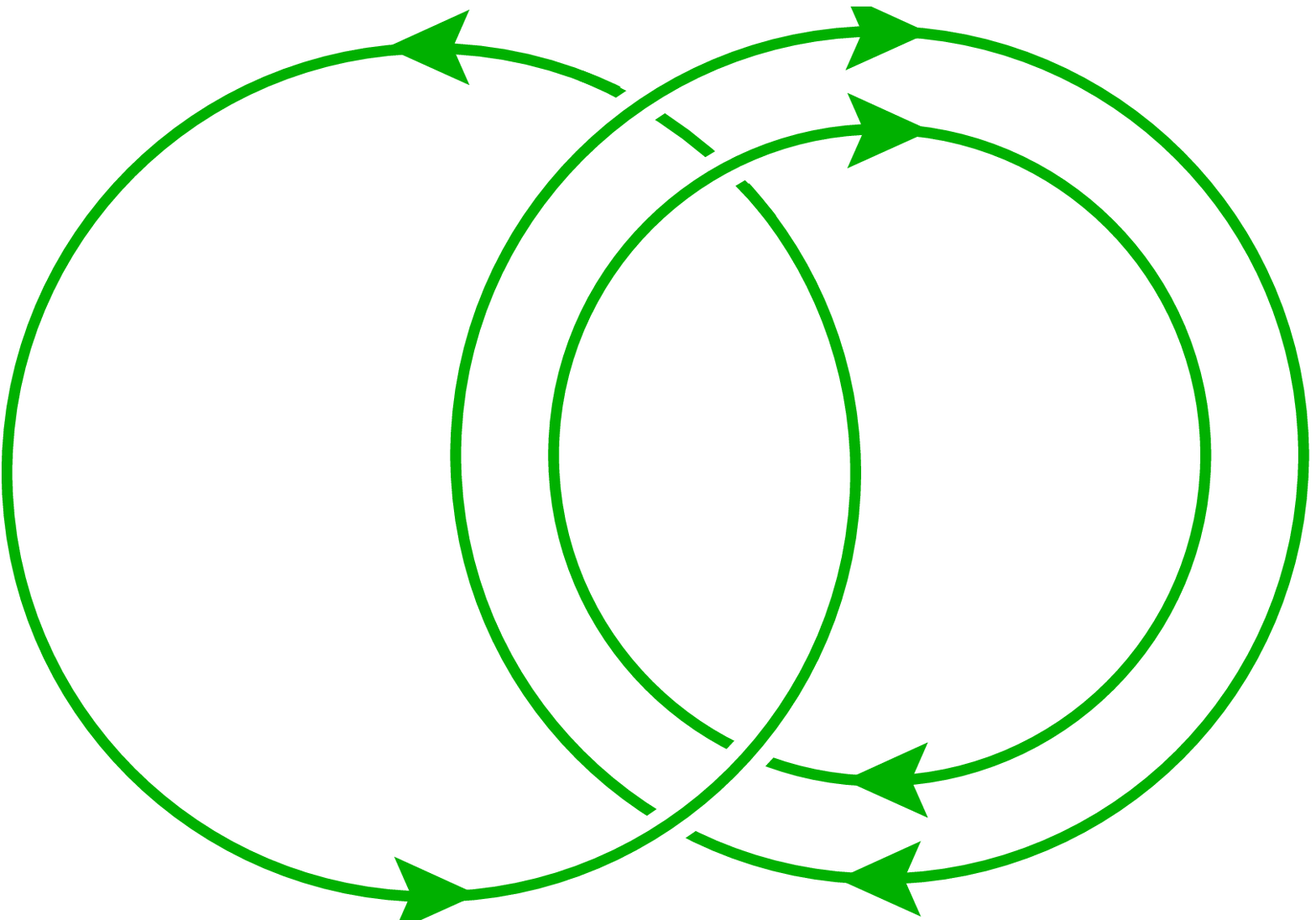}}}
    \put(180,17,5) {$\sss j$}
    \put(189,17.5) {$\sss i$}
    \put(202,17.5) {$\sss k$}
  \put(-22,6){\begin{picture}(0,0)
    \put(88,-55)   {\scalebox{.15}{\includegraphics{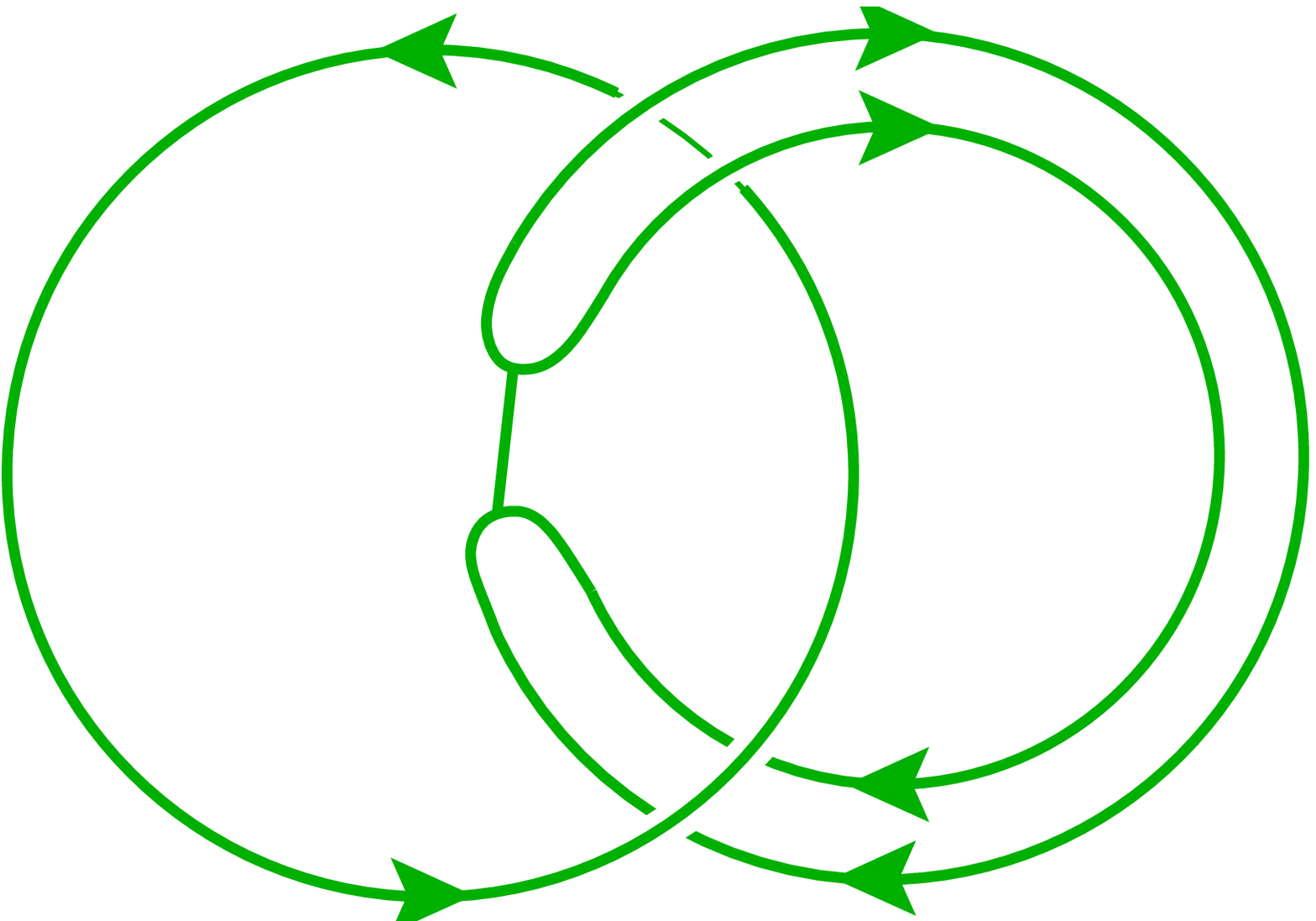}}}
    \put(40,-38)   {$=~~\dsty\sum_p\sum_\alpha$}
    \put(106.2,-35){$\sss p$}
    \put(114.5,-32.4){$\sss \overline\alpha$}
    \put(113.7,-37){$\sss \alpha$}
    \put(127.5,-37){$\sss k$}
    \put(134.5,-32){$\sss i^{\!\vee}$}
    \put(147.5,-32){$\sss j^{\!\vee}$}
    \put(155,-38)  {$=~~\dsty\sum_p\sum_\alpha $}
    \put(200,-55)  {\scalebox{.15}{\includegraphics{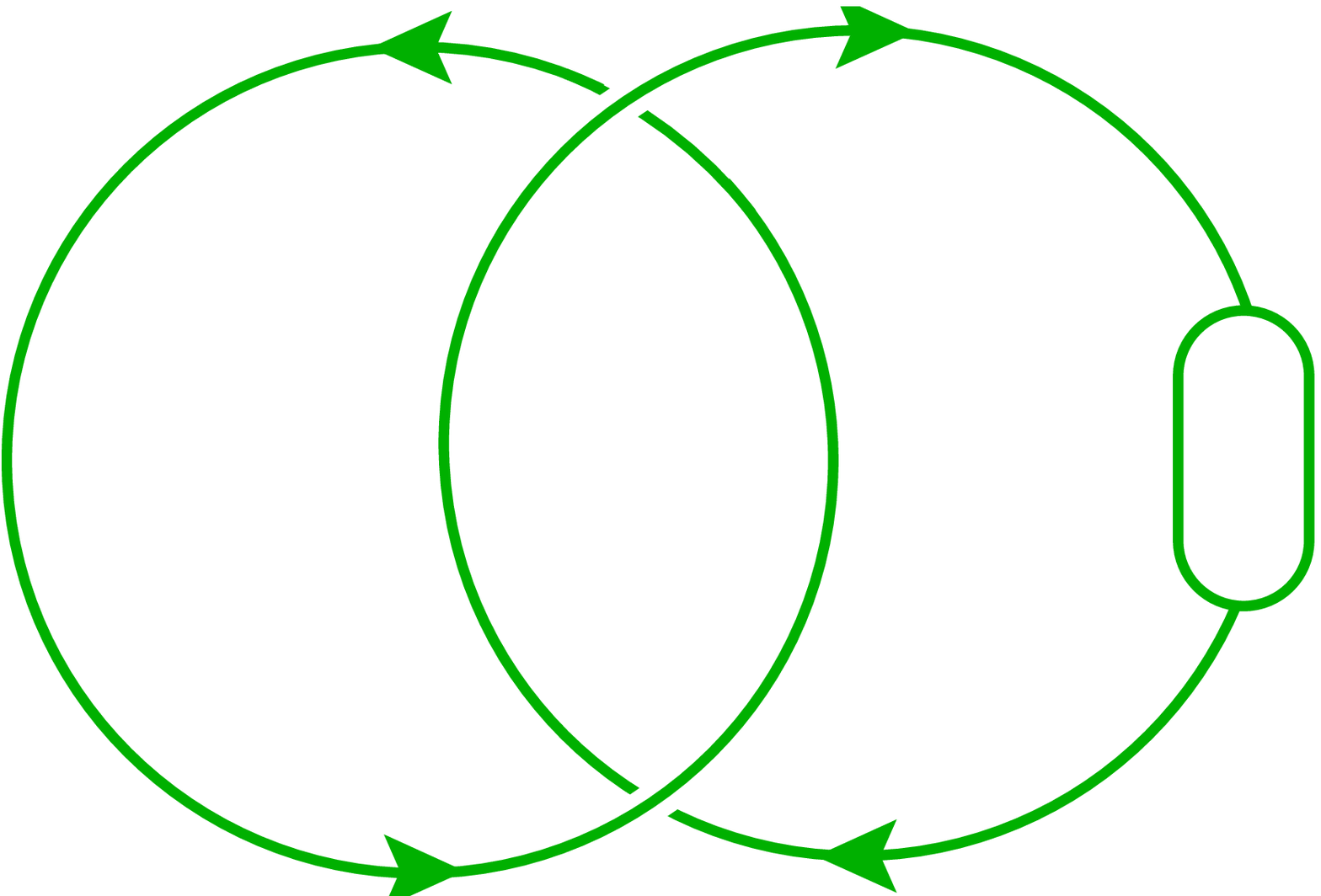}}}
    \put(216,-37)  {$\sss p$}
    \put(233.5,-37){$\sss k$}
    \put(248.5,-34){$\sss i^{\!\vee}$}
    \put(260.5,-34){$\sss j^{\!\vee}$}
    \put(256.8,-46){$\sss \overline\alpha$}
    \put(257.4,-26.8){$\sss \alpha$}
    \put(269,-38)  {$=~~\dsty\sum_p \N ij{\,p} \srO_{p,k}$}
  \end{picture}}
\end{picture}

\noindent
(The first equality is the definition \erf{sr} of $\srO_{j,k}$, the second 
equality uses the same definition for $\srO_{i,k}$, the third is obtained by 
expressing $\id_{U_j\Oti U_i}$ in terms of dual
bases $\{\alpha\}$ of $\Hom(U_j\Oti U_i,U_p)$ and $\{\overline\alpha\}$ of 
$\Hom(U_p,U_j\Oti U_i)$,
the fourth holds by monoidality of the braiding and (co)evaluation morphisms,
and the final equality follows by the property of the bases $\{\alpha\}$ and 
$\{\overline\alpha\}$ to be dual to each other.)
This tells us that the \rS\ \Sr\ diagonalizes the fusion rules, too, and hence 
that it coincides with the \dS\ \erf{Sd}, up to the inherent freedom 
present in \Sd, i.e.\ up to changing the order of the columns (recall the lack 
of a natural bijection between $\Is$ and $\I$) and multiplying them by 
invertible scalars. Thus, in this sense, $\Sr\eq\Sd$, as claimed.

The reasoning above summarizes the proof obtained in 1989, in several different
guises, in \cite{witt27,mose3,card9}. For related material, see e.g.\ 
\cite{mose,diVe,frki,bryz,blth}).
We also note that the equality \erf{SrSd} implies in particular that for any
object $U$, its quantum dimension $\dim(U)\eq\srO_{U,0}\,{\equiv}\,d_U \cir
\tilde b_U$ coincides with its \PF\ dimension.

\smallskip

To prove the equality \erf{SdSm}, on the other hand, requires rather different 
ideas and tools, since the contents of \erf{SdSm} is not purely categorical. 
Early investigations dealt with Virasoro unitary minimal models \cite{besc}. 
Later attention focused on the unitary WZW \cfts\ based on untwisted affine 
Lie algebras $\hat{\mathfrak g}$ and positive level $\ell$, which are 
distinguished in the following two respects: First, there is a convenient 
explicit 
formula expressing the \mS\ \SmOO\ in terms of quantities from the horizontal 
subalgebra $\mathfrak g$ of $\hat{\mathfrak g}$ \cite{kape3}. And secondly,
the sheaves of conformal blocks on a complex curve $C$ can be identified with 
holomorphic sections in the $\ell$th tensor power of a line bundle over the 
moduli space of flat $\mathfrak g$-connections over $C$, so that they may be 
regarded as non-abelian generalizations of theta functions (see e.g.\
\cite{tsuy,beau8,sorg}). Accordingly, the first proofs of \erf{SdSm} for the 
WZW case combined the explicit expression for 
\SmOO\ with al\-geb\-ro-geometric techniques \cite{falt,beLa,beau8}. 
In another proof \cite{tele} the integrable highest weight 
$\hat{\mathfrak g}$-mo\-du\-les are completed to Hilbert spaces and the Verlinde
formula is reduced to vanishing theorems for certain complexes of these Hilbert 
spaces. Yet another \cite{fink} is obtained by establishing a monoidal 
equivalence between the \cat\ of integrable highest weight
$\hat{\mathfrak g}$-mo\-du\-les for a unitary WZW model at level $\ell$ and a 
\cat\ of $\hat{\mathfrak g}$-mo\-du\-les at level $-\ell{-}2h^{\!\vee}$, which 
allows one can invoke the results of \cite{kaluX} for the Grothendieck ring 
of the latter \cat.

Much more recently, the equality $\Sd\eq\SmOO$ has finally been established for 
the \rep\ \cat\ of every \rcva, and in particular without the need to refer to 
any concrete expression for \SmOO\ such as the one available in the WZW case. 
Various pertinent aspects had been clarified 
earlier, e.g.\ in \cite{zhu3,hule3,huan3,hule5,dolm13,miya6}. But the 
proof could be completed, in 2004 \cite{huan20}, only after another crucial 
input, the modular transformation behavior of (genus zero and one) multi-point 
conformal blocks, had been resolved \cite{huan18}.


\section{\Nssi\ fusion rules}\label{sec.nssi}

Let us now study the fusion algebra \F\ for a braided \ftc\ \C\ that is neither 
(necessarily) \ssi\ nor sovereign. \F\ is a \findim\
unital commutative associative \koe-\alg, with a distinguished basis, the fusion
basis $\{[U_i]\,|\,i\iN\I\}$. The structure of such algebras is slightly more 
complicated than in the \ssi\ case, see e.g.\ 
\cite[Chapters\,I.4\,\&\,II.5]{AUrs}.
As a vector space, \F\ can be written as a sum of its Jacobson radical \Rad\
and its \ssi\ part. Thus it has a basis
  \begin{equation}
  \{ \Y_l \,|\, l\iN\Is \} = \{e_a \,|\, a\iN\Ip\} \cup
  \{w_{a,\ell} \,|\, a\iN\Ip\,,\ell\eq1,2,...\,,\nua{-}1 \}
  \label{Ip} \ee
with $\nua\iN\zet_{>0}$ for $a\iN\Ip$, consisting of idempotents $e_a$ and 
nilpotent elements $w_{a,\ell}$
that satisfy the relations $\sum_{a\in\Ip} e_a\eq\one$ and
  $$
  e_a\fus e_b = \delta_{ab}\,e_b \,,\qquad
  w_{a,\ell}\fus w_{b,\ell'} = 0 \quad\mbox{for}~~a\ne b \,,\qquad
  w_{a,\ell}\fus e_b = \delta_{ab}\,w_{a,\ell}
  $$
(the $w_{a,\ell}$ may be chosen such that 
$w_{a,\ell}^{}\eq (w_{a,1})_{}^{*\ell}$), so that
  $$
  \F \,\cong\, \Rad \oplus \bigoplus_{a\in\Ip} \koe e_a  \qquad{\rm with}\qquad
  \Rad \,\cong\, \bigoplus_{a\in\Ip} \bigoplus_{\ell=1}^{\nua-1} \koe w_{a,\ell}
  $$
as \koe-vector spaces.

It follows that in the $\Y_l$-basis, for every $i\iN\I$ the matrix
$M_i\,{:=}\,\Reg([U_i])$ is block-diagonal, with $|\Ip|$ blocks of sizes 
$\nua$. Thus the matrices $N_i$ formed by the structure constants in the 
fusion basis (i.e., having entries $(N_i)_{\!j}^{\,k} \eq \N ijk$), while
in general not diagonalizable, are still {\em block-diagonalizable\/}, 
satisfying $\PP^{-1}_{} N_i\,\PP \eq M_i$ with $\PP$ the matrix for the 
basis transformation from the $\Y_l$-basis to the fusion basis,
$[U_i]\eq\sum_{l\in\Is}\!\PP_{i,l}\,\Y_l$.
Now recall from \erf{Sd} that in the \ssi\ case (for which $\nua\eq1$ for
all $a\iN\Ip\eq\Is$ and all $M_i$ are diagonal), the \dS\ \Sd\ is
obtained from the basis transformation matrix $\PP$ as
$\PP\eq\Sd K$ with $K_{l,m}\eq\delta_{lm}^{}/\SdO_{0,m}$, where the role
of the diagonal factor $K$ is to make $\Sd$ unitary.
This suggests to implement an analogous factorization also in the general 
case, i.e.\ to set
  \begin{equation}
  \PP =: \Sd K \qquad{\rm with}\qquad K = \bigoplus_{a\in\Ip}^{}\! K^a_{}
  \label{SK} \ee 
with $\nua{\times}\nua$-matrices $K^a$. Then one arrives at the equality
  \begin{equation}
  N_i = S \, \KMKi \, S^{-1}_{} \qquad{\rm with}\quad
  S \equiv \SO \quad{\rm and}\quad \KMKi \equiv K\, M_i\, K^{-1}_{}
  \label{Nbd} \ee 
as a generalization of \erf{Nd} to \nssi\ fusion rules.

It should be stressed that it is completely unclear whether any particular 
choice of $K$ is distinguished in applications to \cft. Anyhow,
note that in the \ssi\ case one has $\SdO_{0,l}K_l^{} \eq \PP_{0,l} \eq 1$
for every $l\iN\Is$, so that $K$ is expressible entirely through \Sd.
In contrast, in the general case the condition $\sum_{a}\! e_a\eq\one$
just yields (denoting by $\Ia$ the subset of $\Is$ determined by 
$\{ \Y_l \,|\, l\iN\Ia \} 
\eq \{e_a\}\,{\cup}\,\{w_{a,\ell}\,|\,\ell\eq1,2,...\,,\nua{-}1 \}$)
  $$
  \big(\! \sum_{m\in\Ia}\! \SdO_{0,m}\,K_{m,l}^{a}\, {\big)}_{l\in\Ia}
  = \big(\, \PP_{0,l}\, {\big)}_{l\in\Ia}
  = \big(\, 1 \underbrace{0~\cdots~0}_{\nua-1~\rm times} \!\big)
  $$
for each $a$, i.e.\ only a single restriction on the $\nu_{\!a}^2$ entries of
the block $K^a$.
In addition, by comparison with the \ssi\ case it is tempting to demand   
that the entries of $K^a$ can be expressed solely through the numbers
$\SdO_{0,m}$ with $m\iN\Ia$.
If, for instance, $\nua\eq2$, then this just leaves two parameters in $K^a$
undetermined, one of them being the normalization of $w_a$. Ordering the 
$\Y_l$-basis such that $\Y_\lea^{}\eq e_a$ and $\Y_\lwa^{}\eq w_a$, and --
motivated by the results of \cite{fhst} which imply e.g.\ that \Sd\ should 
square to the unit matrix (see section \ref{sec.epmimo} below), though not by 
any further insight -- fixing these two parameters by imposing that ${\rm det}
\,K^a_{} \eq 1$ and $K^a_{\lea,\lea}\,{+}\,K^a_{\lwa,\lea}\eq0$, one then has 
  \begin{equation}
  \quad K^a_{} \,=\, \left(\!\! \begin{array}{rc}
     \big( \SmS_{0,\lea} \big)_{}^{\!-1} \!\!& -\SdO_{0,\lwa} \\{}\\[-8pt]
  -\,\big( \SmS_{0,\lea} \big)_{}^{\!-1} \!\!&  \SdO_{0,\lea} \eear\!\!\!\right) 
  \qquad{\rm with}\quad \SmS_{0,\lea} := \SdO_{0,\lea}-\SdO_{0,\lwa} \,,
  \label{K} \ee 
and thus
  $$
  \big( \KMKi \big)^{\!a}_{} \!=
  \dsty\frac 1 {\SmS_{0,\lea}}\! \left(\!\!\!  \begin{array}{cc}
  \SdO_{0,\lea}\SdO_{i,\lea}+\SdO_{0,\lwa}\SdO_{i,\lwa}
                            -2\,\SdO_{0,\lwa}\SdO_{i,\lea} \!\!\!\!\!&
  \SdO_{0,\lea}\SdO_{i,\lwa}-\SdO_{0,\lwa}\SdO_{i,\lea} \\{}\\[-6pt]\!\!
  \SdO_{0,\lwa}\SdO_{i,\lea}-\SdO_{0,\lea}\SdO_{i,\lwa} &\!\!\!\!\!\!\!\!\!
  \SdO_{0,\lea}\SdO_{i,\lea}+\SdO_{0,\lwa}\SdO_{i,\lwa}
                            -2\,\SdO_{0,\lea}\SdO_{i,\lwa}
  \eear\!\!\!\right) \!.
  $$

\medskip

Recall that the projective covers $P_i$ of the simple objects $U_i$ close
under the tensor product, see formula \erf B. As a consequence, their images in 
\kc\ span an (associative but, unless \C\ is \ssi, not unital) ideal \P\ of \F.
Together the structure of \F\ and of \P\ already encode a lot of information 
about the tensor product of \C. (Compare the analogous situation with 
tensor products of modules over simple Lie algebras and over quantum groups, 
see e.g.\ \cite{lach,fRSt}.)
If the ideal \P\ is maximal, then it contains in particular the Jacobson radical
\Rad\ of \F, so that the quotient $\F/\P$ is \ssi. While this is indeed the 
case for the \epmo s studied in section \ref{sec.epmimo} below, it is not clear 
to us whether it remains true for other \ftcs.


\section{Modular transformations}\label{sec.mt}

For any \cco\ \cva\ \V\ the space $\Tob\eq\Tob(\V)$ of conformal zero-point 
blocks on the torus carries a \findim\ \rep\ \rx\ of the modular group \slz. 
This has long been known for rational \cva s \cite{zhu3}, for which Zhu's 
algebra \zhuv\ is \ssi. However, \ssy\ of \zhuv\ is not necessary 
\cite[theorem\,5.8]{miya8}. Note that \cco ness
implies \cite[theorem 2.5]{miya8} that \zhuv\
as well as the higher Zhu \alg s \zhuvn\ \cite{dolm9} are \findim.
Further, \cco ness is equivalent to the statement that every weak module is 
\NN-gradable and is a direct sum of \gz eigenspaces to $L_0$ of 
\cite[theorem\,2.7]{miya8}, so that one can define characters as in formula 
\erf{chii} and analogous trace functions for vectors $v\iN\V$ other than the 
vacuum vector, as well as for generalizations known as ``pseudo-trace 
functions". As shown in theorem 5.2 of \cite{miya8}, the space \Tob\ is spanned 
by pseudo-trace functions for the vacuum vector. For rational \V, $\Tob\eq\Chi$
is already spanned by the ordinary characters $\chii_i\,{\equiv}\,\chii_{U_i}$ 
of the irreducible \V-modules $U_i$, whereas in the non-rational case one needs 
in addition nontrivial pseudo-trace functions which are linear combinations of 
characters with coefficients in $\complex[\tau]$ \cite[proposition\,5.9]{miya8}.
(For the \epmimo s these \gz characters coincide \cite{flga} with the functions
introduced on similar grounds in \cite{floh4}.)
There does not, however, exist a canonical assignment of these additional
nontrivial pseudo-traces to particular indecomposable \V-\rep s. (Note that
exactness of a sequence $0\To U\To W\To V\To 0$ implies that
$\chii_W \eq \chii_U \,{+}\, \chii_V $; in particular, the modular 
transformations of any character $\chii_W$ are completely determined by those 
of the irreducible characters.)
As a consequence, the \findim\ \slz-\rep\ \rx\ on the zero-point torus blocks 
does not come with a distinguished basis, unless \V\ is rational, in which 
case a distinguished basis is given by the irreducible characters.

In the \nssi\ case, the proper subspace \Chi\ of \Tob\ is generically 
not invariant under the action of \rt. Rather, the modular transformations
  \begin{equation}
  \chii_i(\gamma\tau) = \sum_{j\in\I} \J_{ij}(\gamma;\tau)\, \chii_j(\tau)
  \label{J} \ee
of the characters involve matrices $\J(\gamma;\tau)$, satisfying 
$\J(\gamma\gamma';\tau) \eq \J(\gamma;\gamma'\tau)\,\J(\gamma';\tau)$
for $\gamma,\gamma'\iN\slz$, which depend nontrivially on $\tau$.  
On the other hand, for there to be any chance to generalize the relation
\erf{SdSm} between the modular S-transformation and the ($K_0$) fusion rules,
it seems indispensible to still have an \slz-\rep\ \rx, of dimension $|\I|$ 
(i.e.\ the dimension of the fusion algebra \F, which equals the number of
irreducible \V-characters), on the subspace \Chi.
Several methods have been proposed for extracting such a \rep\ \rx, and thereby
a candidate \mS\ $\Sm \,{:=}\, \rx(\zzmatrixS)$ from the theory:
\\[2pt]
\nxt~%
Separate \cite{fhst} from the $\tau$-dependent matrices $\J$ a matrix-valued 
automorphy factor \jj\ by writing
  $$
  \J(\gamma;\tau) = \jj(\gamma;\tau)^{-1} \rx(\gamma)_{\phantom|}^{\phantom|} ,
  $$
analogously as separating a scalar automorphy factor
$ \zeta_{c,d}^{-1}(c\tau{+}d)^{-1/2}_{}\,\eE^{-\ii\pi c\nu^2/(c\tau + d)}_{}$
from the Jacobi theta function
$\vartheta(\tau,\nu)$ makes it invariant under $\Gamma^{}_{\!1,2}\,{<}\,\slz$.
Actually, whenever $\jj$ fulfils the cocycle condition
$\jj(\gamma\gamma';\tau) \eq \jj(\gamma';\tau)\,\jj(\gamma;\gamma'\tau)$
and strongly commutes with $\rx\eq\jj\J$, i.e.\ obeys
$[\rx(\gamma),\jj(\gamma';\tau)]\eq0$ for all $\gamma,\gamma'\iN\slz$,
$\rx$ is indeed an \slz-\rep. Thus in order to obtain a sensible result, one
must impose further criteria, e.g.\ suitable minimality conditions, that allow 
one to select an appropriate automorphy factor \jj.
\\[2pt]
\nxt~%
Find \cite{fgst} two strongly commuting \slz-\rep s $\rho_\jj$ and \rxt\ such 
that \rt\ can be written as their ``pointwise product", i.e.
  \begin{equation}
  \rt(\gamma) = \rho_\jj(\gamma) \, \rxt(\gamma)
  \label{rxt} \ee
for $\gamma\iN\slz$, and such that \rxt\ restricts to the \rep\ \rx\ on the
subspace $\Chi\,{\subset}\,\Tob$. 
\\
(The restriction of $\rho_\jj$ to \Chi\ is then essentially the inverse 
automorphy factor $\jj^{-1}$.)
\\[2pt]
\nxt~%
Envoke \cite{fgst} an equivalence between \C\ and the \rep\ \cat\ of a
suitable ribbon quantum group and observe that on the center of the 
quantum group there is an \slz-\rep\ that, via the Jordan decomposition of 
the ribbon element, has a natural pointwise factorization of the form \erf{rxt}. 

Remarkably, these approaches indeed work (and all lead to the same 
\slz-\rep) in the particular case of the \epmimo s, to which we therefore
now turn our attention.


\section{The \epmimo s}\label{sec.epmimo}

\subsection{Representations of the chiral algebra}

The vertex algebra $\V\eq\V_{q,p}$ for the $(q{,}p)$ Virasoro minimal 
model, of central charge $c_{q,p}\eq1\,{-}\,6(p{-}q)^2/pq$, has a \ssi\ \rep\ 
\cat\ iff the positive integers $p$ and $q$ are coprime \cite{wang',domz}.
Here we consider the degenerate case of minimal models with $q\eq1$ 
\cite{kaus}, of central charge $c\eq 13{-}6p{-}6/p$, for which \repV\ is not 
\ssi. For these \epmimo s, \V\ is still \cco, as shown for $p\eq2$ 
(the ``symplectic 
fermion" case \cite{kaus2}) in \cite{abe4}, and for general $p$ in \cite{caFl}.

Note that for $q\eq1$ the Kac table is empty; therefore one considers Virasoro 
modules with labels in the extended Kac table instead. The resulting models 
have an extended chiral algebra \W\ \cite{kaus,floh4}. \W\ can be constructed 
as follows (see e.g.\ \cite[Sect.\,2]{fhst}). Consider a free boson $\varphi$ 
with energy-mom\-en\-tum tensor $\frac12\,\normo\partial\varphi\,
\partial\varphi\normo +\frac12\,(\alpha_+{+}\alpha_-)\,\partial^2\varphi$,
with $\alpha_+\eq\sqrt{2p}$ and $\alpha_-\eq{-}\sqrt{2/p}$.
For any $r,s\iN\zet$ there is a free boson vertex operator $\varPhi_{r,s} \eq
\normo\exp\big(\frac12[(1{-}r)\alpha_+{+}(1{-}s)\alpha_-]\varphi\big)\normo$
and an associated Fock module $F_{r,s}$ over the Heisenberg algebra that is
generated by the modes of $\partial\varphi$. Then \W\ is the maximal local
subalgebra of the algebra spanned by the fields that correspond to the
states in the kernel of the screening charge $S_-\eq \oint\varPhi_{1,-1}$
on the direct sum $\bigoplus_{r\in\zet,s=1,...,p}F_{r,s}$ of Fock modules.
Concretely, \W\ is generated by three Virasoro primary fields of conformal
weight $2p{-}1$, namely $W^- \eq \varPhi_{3,1}$ and $W^0 \eq [S_+,W^-]$,
$W^+ \eq [S_+,W^0]$, where $S_+\eq \oint\varPhi_{-1,1}$.

Further, for every $s\eq1,2,...\,,p$, the direct sum 
$F_s\,{:=}\,\bigoplus_{r\in\zet}F_{r,s}$ of Fock modules
contains two irreducible \W-modules, or more precisely,
  $$
  \mathrm{Ker}\,S_-\big|_{F_s} = \Up_s \oplus \Um_s
  $$
with pairwise nonisomorphic irreducible \W-modules $\Upm_s$.
The modules $\Upm_p$ are projective, while for each $s\eq1,2,...\,,p{-}1$
there exist non-split extensions of $\Upm_s$ by $\Ump_{p-s}$ and by
$\Ump_{p-s}{\oplus}\,\Ump_{p-s}$, and there are in particular 
indecomposable projective modules $\Ppm_s$ which are the projective covers
of $\Upm_s$ and have Jordan-H\"older series corresponding to
  \begin{equation}
  [\Pp_s] = [\Pm_s] = 2\, [\Up_s] + 2\, [\Um_\ps]  
  \label{PU} \ee
in the Grothendieck ring \cite{fhst}. (See also \cite[appendix\,C]{fgst}
and \cite{fgst2} for the corresponding quantum group modules.)
The action of $L_0$ on $\Upm_s$ ($s\,{\ne}\,p$) has size-2 Jordan blocks, 
while $L_0$ acts semisimply on all their proper subquotients. 

\smallskip

We take the \cat\ \C\ to have as objects the projective modules $\Ppm_s$ 
(with $s\eq1,2,...\,,p$, i.e.\ including $\Ppm_p\Cong\Upm_p$) and their
subquotients. By the results of \cite{caFl}, this should be a braided \ftc\
equivalent to a suitable full sub\cat\ 
of the \cat\ of \gz \V-modules \cite{hulz}. Thus in particular \C\ has, 
up to isomorphism, $2p$ simple objects $\Upm_s$, $s\iN\{1,2,...\,,p\}$,
and it consists of $p\,{+}\,1$ linkage classes (minimal full sub\cats\
such that all subquotients of all objects in the class again belong to
the class); $\Up_1$ is the tensor unit.
Two of the linkage classes contain a single
indecomposable projective (indeed simple) object, namely $\Up_p$ and $\Um_p$, 
respectively, while the others contain two nonisomorphic indecomposable
projectives $\Pp_s$ and $\Pm_{p-s}$, as well as two nonisomorphic simple 
objects $\Up_s$ and $\Um_{p-s}$, $s\iN\{1,2,...\,,p{-}1\}$.
The tensor product of \C\ has been studied in detail \cite{gaKa2}
for $p\eq2$, but not for other values of $p$. The results of \cite{gaKa2}
show in particular that for $p\eq2$ the block structure of \F\ reflects
the linkage structure of \C, in the sense that the block sizes $\{ \nua\,|\, 
a\iN\Ip \}$, coincide with the numbers of nonisomorphic simple (or of 
indecomposable projective) objects in the linkage classes of \C. It is tempting
to expect that this continues to be the case for any $p$, so that in
particular the $[\Up_1]$-row of the diagonalizing matrix $\PP$ of \F\ looks as
  \begin{equation}
  \big(\PP_{0,l}\big)_{l\in\Is}^{}
  = \big(\,1~1~\underbrace{1~0~1~0~\cdots~1~0}_{p-1~\rm times}\,\big) .
  \label{10} \ee

\subsection{Characters and modular transformations}\label{subsec.modtr}

As already mentioned at the end of the previous section, one can isolate
an $|\I|\eq2p$\,-dimensional \slz-\rep\ from the modular transformations of
characters by splitting off a suitable automorphy factor from the matrices
$\J(\gamma;\tau)$ that were introduced in \erf J. The characters of the 
\W-modules $\Upm_s$ are linear combinations of classical theta functions 
$\theta_{s,p}(q,z) \eq \sum_{m\in\zet+s/2p} z^m q^{p m^2}$ and their derivatives
(\cite[Proposition\,3.1.1]{fhst}, see also section 6.5 of \cite{floh12}):
  $$ 
  \bearll  \chii_{\Up_s}(q) = \eta(q)^{-1}_{}
  \big(\frac{s}{p}\,\theta_{p{-}s,p}(q) + 2\,\theta'_{p{-}s,p}(q)\big) \,,
  \\{}\\[-4pt]
  \chii_{\Um_s}(q) = \eta(q)^{-1}_{}
  \big(\frac{s}{p}\,\theta_{s,p}(q) - 2\,\theta'_{s,p}(q)\big) \eear
  $$ 
with $\theta_{s,p}(q)\eq\theta_{s,p}(q,1)$ and
$\theta'_{s,p}(q)\eq z\frac\partial{\partial z}\theta_{s,p}(q,z)\!\big|_{z=1}$.
Ordering the $2p$ irreducible characters according to
  $$
  \Up_p\,,\; \Um_p\,,\; \Up_1\,,\; \Um_{p{-}1}\,,\;
  \Up_2\,,\; \Um_{p{-}2}\,,\;\dots\,,\,\Up_{p{-}1}\,,\; \Um_1\,,
  $$
their modular S-transformation takes the same block form as indicated by 
\erf{10} for the diagonalizing matrix $\PP$, namely
  $$
  \J\big( \zzmatrixS;\tau\big) = \begin{pmatrix}
    \RA_{0,0}&\RA_{0,1}&\dots&\RA_{0,p-1}\\[2pt]
    \RA_{1,0}&\RA_{1,1}(\tau)&\dots&\RA_{1,p-1}(\tau)\\ \hdotsfor{4}\\[2pt]
    \,\RA_{p-1,0}&\RA_{p-1,1}(\tau)&\dots&\RA_{p-1,p-1}(\tau)\, \end{pmatrix}
  $$
with $p$ $2{\times}2$\,-blocks
  $$
  \bearll
  \RA_{0,0}= \frac{1}{\sqrt{2p}}\begin{pmatrix}
    \,1&\!1\\ \,1&\!(-1)^p \end{pmatrix} ,
  \qquad&
  \RA_{0,t}= \frac{2}{\sqrt{2p}}\begin{pmatrix}
    1&1\\ (-1)^{p-t}&(-1)^{p-t} \end{pmatrix} ,
  \\{}\\[-.4em]&
  \RA_{s,0}=\frac{1}{\sqrt{2p}} \begin{pmatrix}
    \frac{s}{p}     \!\!&(-1)^{p+s}\frac{s}{p}\\[5pt]
    \frac{p{-}s}{p} \!\!&(-1)^{p+s}\frac{p{-}s}{p}\, \end{pmatrix} ,
  \\{}\\[-.4em]
  \multicolumn2l{
  \RA_{s,t} = (-1)^{p+s+t} \mbox{$\sqrt{\frac2p}$} \begin{pmatrix}
    \frac{s}{p}\cos\pi\frac{st}{p}
    -\ii\tau\,\frac{p{-}t}{p}\sin\pi\frac{st}{p}
    \!\!&\frac{s}{p}\cos\pi\frac{st}{p}
    +\ii\tau\,\frac{t}{p}\sin\pi\frac{st}{p}\\[7pt]
    \frac{p{-}s}{p}\cos\pi\frac{st}{p}
    +\ii\tau\,\frac{p{-}t}{p}\sin\pi\frac{st}{p}
    \!\!&\frac{p{-}s}{p}\cos\pi\frac{st}{p}
    -\ii\tau\,\frac{t}{p}\sin\pi\frac{st}{p} \end{pmatrix} .
  }\eear
  $$
We now split off an automorphy factor $\jj$ that fulfils the following
additional criteria: First, it is block-diagonal, with block structure 
reflecting the one of $\J$; and second, it is minimal in the sense that when 
expressed in the basis of \Chi\ that is furnished by the theta
functions and their derivatives, it essentially reduces to the `trivial' 
scalar automorphy factor $(c\tau{+}d)^{-1}$. Concretely, we set
\cite[Proposition\,4.3.1]{fhst}
  \begin{equation}
  \jj(\gamma;\tau) = \onematrix_{2\times2}^{} \,\oplus\, 
  \RB_1(\gamma;\tau) \,\oplus\,\cdots\,\oplus\, \RB_{p-1}(\gamma;\tau)
  \label{j1j} \ee
where, for $s\iN\{1,2,...\,,p{-}1\}$, 
  $$
  \RB_s(\gamma,\tau) = L_s \begin{pmatrix}\,1\!\!&0\\
  \,0\!\!&\zeta(\gamma)\,\alpha(\gamma,\tau) \end{pmatrix} L_s^{-1} 
  \qquad{\rm with}\quad\
  \alpha \big(\zzmatrixabcd;\tau)\big) = \frac1{c\tau+d}
  $$
and $L_s$ the similarity transformation between the two bases
$\{ \chii_{\Up_s},\chii_{\Um_{p{-}s}} \}$ and 
$\{ \theta_{s,p}^{},\theta'_{\!\!s,p} \}$ of the two-dimensional subspace
of \Chi\ spanned by the characters of $\Up_s$ and $\Um_{p{-}s}$,
and with $\zeta$ the \slz-character defined by
$\zeta\big(\zzmatrixS\big)\eq\ii$ and $\zeta\big(\zzmatrixT\big)\eq \kappa$
with $\kappa^3\eq{-}\ii$.
With this automorphy factor one obtains
  $$
  \bearll  \Sm \!\equiv \rx\big(\zzmatrixS\big)\!\!\!\!&
  := \jj\big(\zzmatrixS;\tau\big) \, \J\big(\zzmatrixS;\tau\big)
  \\{}\\[-8pt] &
  = \J\big(\zzmatrixS;\tau{=}\ii\,\big) \,
  = \begin{pmatrix} \RA_{0,0}&\RA_{0,1}&\dots&\RA_{0,p-1}\\[2pt]
    \RA_{1,0}&\rmS_{1,1}&\dots&\rmS_{1,p-1}\\ \hdotsfor{4}\\[2pt]
    \,\RA_{p-1,0}&\rmS_{p-1,1}&\dots&\rmS_{p-1,p-1}\, \end{pmatrix} \eear
  $$
with
  $$
  \rmS_{s,t}=(-1)^{p+s+t} \mbox{$\sqrt{\frac2p}$}\, \begin{pmatrix}
    \frac{s}{p}\cos\pi\frac{st}{p} +\frac{p{-}t}{p}\sin\pi\frac{st}{p}
    &\frac{s}{p}\cos\pi\frac{st}{p} -\frac{t}{p}\sin\pi\frac{st}{p}\\[7pt]
    \,\frac{p{-}s}{p}\cos\pi\frac{st}{p} -\frac{p{-}t}{p}\sin\pi\frac{st}{p}
    &\frac{p{-}s}{p}\cos\pi\frac{st}{p} +\frac{t}{p}\sin\pi\frac{st}{p}\,
  \end{pmatrix} .
  $$
Note that the matrix \Sm\ defined this way squares to the unit matrix 
$\onematrix_{2p\times 2p}$, but that it
is not symmetric. Furthermore, it satisfies 
  \begin{equation}
  \SmO_{2p-m+3,2p-n+3} \eq (-1)
  ^{p(1-\delta_{m,1}-\delta_{n,1})+\lfloor(m+n+1)/2\rfloor+mn}_{}\SmO_{m,n}.
  \label{ssc} \ee

\smallskip
\noindent
Explicitly, we have e.g.
  $$
  \Sm = \begin{pmatrix}
      \,\,{1}/{2} & {1}/{2} & 1 & 1 \\[1pt]
      \,\,{1}/{2} & {1}/{2} & -1 & -1 \\[1pt]
      \,\,{1}/{4} & - {1}/{4} & {1}/{2} & - {1}/{2} \\[1pt]
      \,\,{1}/{4} & - {1}/{4} & - {1}/{2} & {1}/{2}
    \end{pmatrix}
  $$
for $p\eq2$ and 
  $$
  \Sm = \addtolength{\arraycolsep}{-2pt} \begin{pmatrix}
      \,{1}/{{\sqrt{6}}} & {1}/{{\sqrt{6}}} &
        {\sqrt{{2}/{3}}} & { \sqrt{{2}/{3}}} &
        {\sqrt{{2}/{3}}} & {\sqrt{{2}/{3}}} \\[5pt]
      \,{1}/{{\sqrt{6}}} & - {1}/{{\sqrt{6}}} &
        {\sqrt{{2}/{3}}} & {\sqrt{{2}/{3}}} &
        -{\sqrt {{2}/{3}}} & -{\sqrt{{2}/{3}}} \\[4pt]
        \frac{1}{3{\sqrt{6}}} & \frac{1}{3{\sqrt{6}}} & \frac{-( 6 +
          {\sqrt{3}} ) }{9{\sqrt{2}}} & \frac{ 3 - {\sqrt{3}} }
        {9{\sqrt{2}}} & \frac{ 3 - {\sqrt{3}} }{9{\sqrt{2}}} &
        \frac{-( 6 + {\sqrt{3}} ) } {9{\sqrt{2}}} \\[5pt]
        \frac{{\sqrt{2/3}}}{3} & \frac{{\sqrt{2/3}}} {3} & \frac{
          {\sqrt{2}} ( 3 - {\sqrt{3}} ) } {9} & \frac{-( 3 +
          2{\sqrt{3}} ) }{9{\sqrt{2}}} & \frac{-( 3 + 2{\sqrt{3}} )
        }{9{\sqrt{2}}} & \frac{ {\sqrt{2}} ( 3 - {\sqrt{3}} ) } {9} \\[3pt]
        \frac{{\sqrt{2/3}}} {3} & \frac{-{\sqrt{2/3}}}{3} & \frac{
          {\sqrt{2}} ( 3 - {\sqrt{3}} ) } {9} & \frac{-( 3 +
          2{\sqrt{3}} ) }{9{\sqrt{2}}} & \frac{3 +
          2{\sqrt{3}}}{9{\sqrt{2}}} & \frac{ {\sqrt{2}}( {\sqrt{3}} - 3 ) }
        {9} \\[3pt]
        \frac{1}{3{\sqrt{6}}} & \frac{-1} {3{\sqrt{6}}} & \frac{-( 6 +
          {\sqrt{3}} ) }{9{\sqrt{2}}} & \frac{ 3 - {\sqrt{3}} }
        {9{\sqrt{2}}} & \frac{{\sqrt{3}} - 3} {9{\sqrt{2}}} & \frac{6
          + {\sqrt{3}}} {9{\sqrt{2}}}
    \end{pmatrix}
  $$
for $p\eq3$.

\smallskip

The \slz-\rep\ \rt\ on the space \Tob\ of conformal zero-point blocks on the 
torus is $(3p{-}1)$-dimensional \cite{floh4,miya8,flga}.
According to Theorem 2.3 of \cite{fgst} it can be written as a pointwise 
product of the form \erf{rxt}, with the restriction of $\rho_\jj$ to the 
$2p$-dimensional subspace \Chi\ spanned by the characters yielding the inverse 
of the automorphy factor \erf{j1j} and \rxt\ restricting to the \slz-\rep\ \rx.

\subsection{Conjecture \cite{fhst}:
           \boldmath{$K_0(\C)$} from a \gz Verlinde relation}
$\,$\\[-10pt]
We now {\em postulate\/} that the equality \erf{SdSm} holds for the \epmimo s,
with 
\\[1.5pt]
\nxt~\Sd\ as defined in \erf{SK}, subject to the assumption that the block
structure of \F\ is the one indicated in \erf{10},
and to the choice of normalizations made in \erf K, and
\\[-.5pt]
\nxt~\Sm\ given by $\rx\big(\zzmatrixS\big)$ as obtained above.

\smallskip

This is, admittedly, a bold assumption. However, it is at least partly 
justified by the fact that it survives the following
quite non-trivial consistency check: It generalizes the Verlinde formula
\erf v in the sense that the numbers obtained when inserting $S\eq\SmOO$ into
the formula \erf{SK} for the block-\dS, which are priori general complex 
numbers (actually, algebraic numbers in a suitable finite abelian extension of 
the rationals), turn out to be nonnegative integers, as is needed for the 
structure constants of \F\ in the fusion basis. 

Indeed, performing this exercise yields the following unital associative ring 
structure, which is conjectured to coincide with the Grothendieck ring \kc\ 
that results from the tensor product of \C\ \cite{fhst}: 
$[\Up_1]$ is the unit element, while $[\Um_1]$ is an order-$2$ simple current 
(invertible element) acting fixed-point free, such that
  $$
  [\Um_1] \fus [\Upm_s] = [\Ump_s] \,,
  $$
and thereby the remaining relations all reduce to
  \begin{equation}
  [\Upm_s] \fus [\Upm_t] \,=\, [\Um_1] \fus \big(\, [\Upm_s]\fus[\Ump_t] \,\big)
  \,= \sum_{i=1}^t\, [\widehat{\Up_{\phantom|}}{}_{\!\!\!\!\!\!s-t+2i-1}^{}] 
  \label{kc} \ee
for $1\,{\le}\,t\,{\le}\,s\,{\le}\,p$, where
  $$
  [\widehat{\Upm_{\phantom|}}{}_{\!\!\!\!\!\!s}^{}] = \left\{ \bearll
  [\Upm_s] & {\rm for}\quad 1\,{\le}\,s\,{\le}\,p \,, \\[3pt]
  [\Upm_{2p-s}] +2\,[\Ump_{s-p}] & {\rm for}\quad p{+}1\,{\le}\,s\,{\le}\,2p{-}1
  \,.  \eear \right.
  $$ 
It follows in particular that the \PF\ dimensions of the simple objects are 
given by
  $$
  \pf(\Upm_s) = s \qquad{\rm for}\quad s\eq1,2\,...\,,p \,.
  $$
Basis-independently, the structure of \kc\ obtained this way can be described 
\cite[Proposition\,3.3.7]{fgst} as the quotient of the polynomial ring 
$\complex[x]$ by an ideal generated by a certain linear combination of Chebyshev
polynomials, with the variable $x$ corresponding to $[\Up_2]$.

\smallskip

\underline{To summarize:}~~~~(1)~~~The following strategy affords a 
generalization of the Verlinde conjecture to \nssi\ fusion rules:
\\[3pt]
\nxt~Parametrize a suitably normalized matrix \Sd\ that block-diagonalizes \kc.
\\
\nxt~Extract an $|\I|{\times}|\I|$-matrix $\Sm\,{:=}\,\rx(\zzmatrixS)$ from the
data of the CFT.
\\[-1pt]
\nxt~Insert \Sm\ for $S \,{\equiv}\, \Sd$ in the formula \erf{Nbd}, i.e.\
compute $N_i \eq \Sm\, \widehat{\SmO_i}\, {\Sm}^{-1}_{}$.
\\[3pt]
(2)~~~For the \epmimo s, this strategy produces sensible fusion rules.

\medskip

It is also worth recalling that the images of the projective objects of \C\ 
form an ideal \P\ in \kc. The restriction of the fusion product to \P\
is rather degenerate (compare \cite[Section\,3.3.2]{fgst}); e.g.\ for $p\eq2$,
the relations for the three basis elements $[\Up_2]$, $[\Um_2]$ and
$[P_1^{}]\,{:=}\,[\Pp_1]\eq[\Pm_2]$ of \P\ read\,%
  \footnote{~These fusion rules were actually already observed in 
  \cite[Section\,5.3]{rohs}.}
  $$
  \bearll  [\Upm_2] \fus [\Upm_2] = [\Up_2] \fus [\Um_2] = [P_1^{}] \,,\
  \\{}\\[-5pt]
  [\Upm_2] \fus [P_1^{}] = 2\, \big( [\Up_2] + [\Um_2] \big) \,,&
  [P_1^{}] \fus [P_1^{}] = 4\, [P_1^{}] \,. \eear
  $$
But the quotient $\F/\P$ turns out to have a nice structure 
\cite[Corollary\,3.3.5]{fgst}: it is \ssi\ (implying that \P\ contains the
Jacobson radical of \F), and is in fact isomorphic to the fusion ring of the 
$\mathfrak{sl}(2)$ WZW theory at level $p{-}2$.
To see this note that, owing to \erf{PU}, in $\F/\P$ we have $[\Up_s]\,{+}\,
[\Um_{p-s}]\simeq0$ for $s\eq1,2\,...\,,p{-1}$. As a consequence,
  $$
  [\widehat{\Upm_{\phantom|}}{}_{\!\!\!\!\!\!s}\,\,\,] \,\simeq
  {-}\,[\Upm_{2p-s}] \qquad{\rm for}\quad s\eq p{+}1,p{+}2,...\,,2p \,,
  $$
so that $\F/\P$ is spanned by (the images of) $[\Up_s]$ with 
$s\eq1,2\,...\,,p{-1}$, and
  $$
  [\Up_s] \fus [\Up_t] \,\simeq\, \sum_
  {\stackrel{\scriptstyle r=|s-t|+1}{r+s+t\,\in\,2\zet}}^{p-1-|p-s-t|}\! \Up_r
  $$
in $\F/\P$.

\subsection{Relation with the quantum group \Uqres}

By inspection, the \PF\ dimensions $\pf(\Upm_s)$ of the simple objects of \C\
coincide with the dimensions of the simple modules over the
$p$-restricted enveloping algebra of $\mathfrak{sl}(2,\mathbb F_p)$
(see e.g.\ \cite{hump3}); moreover, the same holds for their respective
projective covers, for which $\pf(\Ppm_s)\eq2p$ for all $s\eq1,2,...\,,p$.
In view of the intimate relationship between modular \rep s and quantum groups 
at roots of unity (see e.g.\ \cite{soer5}), this may be taken as an indication 
that there might exist a suitable quantum group with a \rep\ \cat\ equivalent
to the \cat\ \C.

As advocated in \cite{fgst,fgst2}, and proven for $p\eq2$ in \cite{fgst2}, such 
a quantum group indeed exists,, namely the restricted (non-quasitriangular) 
Hopf algebra \Uqres\ with 
the value ${\mathrm q}\eq\eE^{\pi\ii/p}$ of the deformation parameter.
\Uqres\ has $2p$ irreducible modules, and \cite[Theorem\,3.3.1]{fgst} its 
Grothendieck ring is isomorphic, as a fusion ring (i.e.\ via an isomorphism 
preserving distinguished bases), with the conjectured result \erf{kc} for \kc. 

The center of the $2p^3$-dimensional algebra \Uqres\ has dimension $3p{-}1$;
as discussed in \cite{kerl2,lyub6,fgst}, it carries a \rep\ of \slz.
This \rep\ can be shown to be isomorphic to the 
\slz-\rep\ \rt\ on the space of torus zero-point blocks of the \epmimo, and
in terms of \Uqres, the pointwise factorization of \rt\ mentioned at the
end of section \ref{subsec.modtr} can be understood through the Jordan 
decomposition of the ribbon element of \Uqres\ (which essentially gives the
\rep\ matrix in \rt\ for the $T$-transformation) 
\cite[Theorems\,5.2\,\&\,5.3.3]{fgst}.


\section{Outlook}\label{sec.o}

Obviously, the approach taken above is too narrow to apply to general 
non-rational CFTs. Indeed, the non-rational models that fit in our framework 
still share many features of the rational ones. Among the types of 
non-rational CFTs that are excluded are, for instance, Liouville theory
(see e.g.\ \cite{naKA'} for a review, and \cite{jeTr} for the discussion of 
a Verlinde-like relation in a subsector of the theory), all CFTs having
chiral sectors of infinite Perron\hy Frobenius dimension, like the
non-rational orbifolds of a free boson, as well as other interesting 
$c\eq1$ theories such as the $(p{,}p)$ minimal models \cite{mila}.
Nevertheless it seems worth continuing the investigation of (braided) \ftcs\
and their fusion rules. First, one may hope that even for more general
non-rational CFTs there exists a suitable subcategory, e.g.\ the full 
subcategory formed by all objects of finite length, that has the structure
of a braided \ftc\ and still captures interesting features of the CFT.
Secondly, there are several classes of non-rational models which potentially 
do fit into the framework, like other logarithmically extended 
$(p{,}q)$ minimal models, fractio\-nal-le\-vel WZW models (see e.g.\ 
\cite{fema3,gabe10,lmrs,adaM5}), or \gz symplectic fermions \cite{abe4}.

\smallskip

It will be important to determine whether the chiral sectors of models like 
those just mentioned indeed form \ftcs, and furthermore, to which extent 
they share additional features that are present for the \epmimo s. The most 
important of these is the fact that for the \epmo s the linkage structure of 
\C\ exactly matches the linkage structure of the \rep\ \cat\ $\Rep(\F)$ of 
the fusion \alg: every simple \F-module appears as a composition factor of 
precisely one block of \F, and all composition factors of any indecomposable 
\F-module occur in one and the same block.
Some particular aspects of this property of the \epmo s are the following: 
\\[2pt]
\nxt~%
The quotient \alg\ $\F/\P$ is \ssi.
\\[2pt]
\nxt~%
There exist projective simple objects.
(They appear to be analogues of the Stein\-berg modules in the \rep\ theory
of simple Lie algebras or quantum groups.)
\\[2pt]
\nxt~%
If indecomposable projective objects of \C\ belong to the same linkage class, 
then they have the same image in \kc.  (This behavior is familiar e.g.\ from 
certain induced modules over reduced enveloping algebras in the theory of 
modular \rep s \cite{jant6}.) Thus while there are as many isomorphism 
classes of indecomposable projectives as of simple objects,
so that a priori the dimension of \P\ could be as large as $|\I|$, the dimension
of \P\ is actually given by the number of blocks of \kc\ (i.e., in the notation
of \erf{Ip}, by $|\Ip|$).

\smallskip

Another feature of the fusion rules of the \epmimo s is the presence of the
simple current $[\Um_1]$. One would expect that, like in the rational case 
\cite{scya3,intr}, simple currents are accompanied by simple current symmetries 
of the matrix \SmOO; for the \epmo s, this is indeed the case: 
relation \erf{ssc} is a simple current symmetry.\,%
  \footnote{~%
  The Galois group of the finite abelian extension of $\mathbb Q$ in which
  (upon suitable normalization of the nilpotent basis elements and of the
  entries of the matrix $K$) the entries of the block-diagonalizing matrix \Sd\
  take their values gives rise to another type of symmetry. 
  One may hope that these `Galois symmetries' can be
  exploited in an similar manner as \cite{coga,fgss,fusS2,bant16} for rational
  CFTs; in a related context this issue has been addressed in \cite{gann21}.} 
Also worth being investigated is the question whether based solely on the \cat\
\C\ of chiral sectors, or even just on its fusion ring \kc, one can make any
interesting predictions about the Zhu \alg s \zhuvn\ and their \rep s. 
These could then be tested in cases for which the vertex algebra is 
sufficiently well under control. For instance, for the \epmo s one expects 
\cite{flga} that \zhuv\ has dimension $6p{-}1$.  

\smallskip

Finally, we would like to point out that the discussion of CFT above is 
incomplete in that it is exclusively concerned with chiral issues. In most 
applications it is full, local, CFT rather than chiral CFT that is relevant.
How to deal with non-rational full CFT is not understood generally. Only a few 
peculiar results have been obtained, e.g.\ in \cite{gaKa3}, where a local theory
for the $(1{,}2)$ minimal model was constructed, or in \cite{fhst}, where the 
existence of some non-diagonal modular invariant combinations of characters 
was noticed for the $(1{,}2)$ and $(1{,}3)$ models.

An indication of what is going on in the general case is supplied by the
observation that for any full CFT there should be sensible notions of 
topological defect lines and of boundary conditions. These structures play a 
central role in the construction of (rational) full CFTs based on noncommutative
algebra in tensor categories \cite{fuRs4,fuRs11,scfr} (for related work in the 
context of vertex algebras see \cite{kong,huko2}). Indeed, since defect lines 
can be fused with each other, one expects that they form a monoidal \cat\ 
$\mathcal D$, and that the defects can be deformed locally may be taken as an 
indication that $\mathcal D$ should be rigid. Moreover, since a defect 
line can be fused with a boundary condition, resulting in another boundary 
condition, the boundary conditions should form a (left, say) module \cat\ 
$\mathcal M$ over $\mathcal D$. (For various pertinent aspects of module \cats\ 
see \cite{ostr,etno,etos,anFe}.) And, as pointed out in \cite{scfr}, this 
structure gives rise to a formulation in which $\mathcal M$ and $\mathcal D$ 
have naturally the structure of a (right) module \cat\ and a bimodule \cat, 
respectively, over the monoidal \cat\ \C\ of chiral sectors of the CFT. 
Furthermore, from the viewpoint of \cite{scfr} \C\ is actually a derived 
concept, the primary structure being the \cat\ $\mathcal D$, which already in 
rational CFT generically neither has a braiding nor a twist. It is tempting 
to try to analyze also (a subclass of) non-rational full CFTs by starting
from the defect line \cat\ $\mathcal D$, taken to be a \ftc. More generally, 
the existence of module and bimodule \cats\ with matching properties
might serve as a guiding principle when trying to find necessary
properties that a monoidal \cat\ must possess in order to correspond to 
a non-rational chiral CFT.

\bigskip

\noindent {\small
{\sc Acknowledgements:}\\ I am grateful to Matthias Gaberdiel, Terry Gannon, 
Christoph Schweigert and Aliosha Semikhatov for helpful discussions and 
comments on the manuscript.}
\bigskip


 \newcommand\wb{\,\linebreak[0]} \def\wB {$\,$\wb}
 \newcommand\Bi[2]    {\bibitem[#2]{#1}}
 \newcommand\JO[6]    {{\em #6}, {#1} {#2} ({#3}), {#4--#5} }
 \renewcommand\J[7]     {{\em #7}, {#1} {#2} ({#3}), {#4--#5} {{\tt \,[#6]}}}
 \newcommand\JJ[4]    {{\em #4}, {#1} {#2} ({#3})}
 \newcommand\Phd[2]   {{\em #2}, Ph.D.\ thesis (#1)}
 \newcommand\PhD[3]   {{\em #2}, Ph.D.\ thesis (#1) {\tt \,[#3]}}
 \newcommand\Pret[2]  {{\em #2}, pre\-print {\tt #1}}
 \newcommand\BOOK[4]  {{\em #1\/} ({#2}, {#3} {#4})}
 \newcommand\inBO[8]{{\em #8}, in:\ {\em #1}, {#2}\ ({#3}, {#4} {#5}), p.\ {#6--#7}}
 \def\dim   {dimension}
 \def\jf    {J.\ Fuchs}
 \def\adma  {Adv.\wb Math.}
 \def\anma  {Ann.\wb Math.}
 \def\aspm  {Adv.\wb Stu\-dies\wB in\wB Pure\wB Math.}
 \def\aste  {Ast\'e\-ris\-que}
 \def\bams  {Bull.\wb Amer.\wb Math.\wb Soc.}
 \def\cocm  {Com\-mun.\wb Con\-temp.\wb Math.}
 \def\coma  {Con\-temp.\wb Math.}
 \def\Coma  {Con\-temp. Math.}
 \def\comp  {Com\-mun.\wb Math.\wb Phys.}
 \def\Comp  {Com\-mun.\wb Math. Phys.}
 \def\cpma  {Com\-pos.\wb Math.}
 \def\duke  {Duke\wB Math.\wb J.}
 \def\dukE  {Duke Math.\wb J.}
 \def\fiic  {Fields\wB Institute\wB Commun.}
 \def\Fiic  {Fields Institute\wB Commun.}
 \def\fiiC  {Fields\wB Institute Commun.}
 \def\foph  {Fortschritte\wb Phys.}
 \def\gafa  {Geom.\wB and\wB Funct.\wb Anal.}
 \def\ihes  {Publ.\wb Math.\wB I.H.E.S.}   
 \def\imrn  {In\-tern.\wb Math.\wb Res.\wb No\-ti\-ces}
 \def\inma  {Invent.\wb math.}
 \def\ijmp  {Int.\wb J.\wb Mod.\wb Phys.\ A}
 \def\jams  {J.\wb Amer.\wb Math.\wb Soc.}
 \def\jamS  {J.\wb Amer.\wb Math. Soc.}
 \def\jhep  {J.\wb High\wB Energy\wB Phys.}
 \def\joac  {J.\wB Al\-ge\-bra\-ic\wB Com\-bin.}
 \def\joag  {J.\wB Al\-ge\-bra\-ic\wB Geom.}
 \def\joal  {J.\wB Al\-ge\-bra}
 \def\jomp  {J.\wb Math.\wb Phys.}
 \def\jopa  {J.\wb Phys.\ A}
 \def\josp  {J.\wb Stat.\wb Phys.}
 \def\jpaa  {J.\wB Pure\wB Appl.\wb Alg.}
 \def\mams  {Memoirs\wB Amer.\wb Math.\wb Soc.}
 \def\momj  {Mos\-cow\wB Math.\wb J.}
 \newcommand\ncmp[3] {\inBO{IXth International Congress on
            Mathematical Physics} {B.\ Simon, A.\ Truman, and I.M.\ Da\-vis, 
            eds.} {Adam Hilger}{Bristol}{1989} {#1}{#2}{#3} } 
 \def\npbp  {Nucl.\wb Phys.\ B (Proc.\wb Suppl.)}
 \def\nupb  {Nucl.\wb Phys.\ B}
 \def\Nupb  {Nucl. Phys.\ B}
 \def\phlb  {Phys.\wb Lett.\ B}
 \def\phrl  {Phys.\wb Rev.\wb Lett.}
 \def\pnas  {Proc.\wb Natl.\wb Acad.\wb Sci.\wb USA}
 \def\pspm  {Proc.\wb Symp.\wB Pure\wB Math.}
 \def\reth  {Represent.\wB Theory}
 \def\rims  {Publ.\wB RIMS}
 \def\rvmp  {Rev.\wb Math.\wb Phys.}
 \def\slnm  {Sprin\-ger\wB Lecture\wB Notes\wB in\wB Mathematics}
 \newcommand\Slnm[1] {{\rm[\slnm\ #1]}}
 \def\trgr  {Trans\-form. Groups}
   \def\AMS    {{American Mathematical Society}}
   \def\AW     {{Addi\-son\hy Wes\-ley}}
   \def\BIR    {{Birk\-h\"au\-ser}}
   \def\Be     {{Berlin}}
   \def\Bo     {{Boston}}
   \def\Ca     {{Cambridge}}
   \def\CUP    {{Cambridge University Press}}
   \def\MD     {{Marcel Dekker}}
   \def\NY     {{New York}} 
   \def\OUP    {{Oxford University Press}}
   \def\PR     {{Providence}}
   \def\SV     {{Sprin\-ger Ver\-lag}}
   \def\WS     {{World Scientific}}
   \def\Si     {{Singapore}} 

\bibliographystyle{amsalpha}  \medskip
\end{document}